\documentclass[a4paper,12pt]{article}
\usepackage{epsfig}
\usepackage{doublespace}
\usepackage{subfigure}

\def\slashchar#1{\setbox0=\hbox{$#1$}           % set a box for #1
   \dimen0=\wd0                                 % and get its size
   \setbox1=\hbox{/} \dimen1=\wd1               % get size of /
   \ifdim\dimen0>\dimen1                        % #1 is bigger
      \rlap{\hbox to \dimen0{\hfil/\hfil}}#1 
   \else                                        % / is bigger
      \rlap{\hbox to \dimen1{\hfil$#1$\hfil}}/                                    \fi}          

\newcommand\pmiss{{{\slashchar{p}}}}

\newcommand\Ptmiss{\slashchar{{\bf p}}_T}

\def\ie{i.e.}
\def\eg{e.g.}
\def\ptwo{{\bf p}}

\def\PtlepOne{{{\bf p}_T^{l_1}}}
\def\PtlepTwo{{{\bf p}_T^{l_2}}}
\def\Ptalpha{{{\bf p}_T^{\alpha}}}
\def\Ptbeta{{{\bf p}_T^{\beta}}}
\def\mttwo{{m_{T2}}}
\def\mttwomax{{m_{T2}^{\rm{max}}}}
\def\chginooneplus{{\chi^+_1}}
\def\chginoonepm{{\chi^\pm_1}}

\def\ntlinoone{{\chi^0_1}}

\def\ntlone{\ntlinoone}
\def\chgone{\chginooneplus}
\def\mtthree{{m_{T3}}}
\def\mtfour{{m_{T4}}}
\def\mttwosq{{m_{T2}^2}}

\def\half{{\frac 1 2}}

\def\mynabla{{\Delta}}
\def\mybigq{{Q}}
\def\tbar{{\bar t}}
\def\ttbar{{t \tbar}}
\def\ifb{{{\mathrm {fb}}^{-1}}}
\def\GeV{{{\mathrm {GeV}}}} 
\newcommand{\slptwo}{{{\slashchar{{\bf p}}}}}

\newcommand{\smallTwoVec}[2]{{\renewcommand\arraystretch{0.6} \begin{array}[c]{c} 
\!\!\!    #1    \!\!\!\!
\\ 
\!\!\!    #2    \!\!\!\!
\end{array} }}

\newcommand{\smallThreeVec}[3]{{\renewcommand\arraystretch{0.6} \begin{array}[c]{c} 
\!\!\!    #1    \!\!\!\!
\\ 
\!\!\!    #2    \!\!\!\!
\\ 
\!\!\!    #3    \!\!\!\!
\end{array} }}

\newcommand{\capbox}[2]{\parbox{0.85\textwidth}{\caption[#1]{\textit{#2}}}}

\def\susic{supersymmetric}

\newcommand{\mysecref}[1]{section~\ref{#1}}
\newcommand{\myeqref}[1]{(\ref{#1})}
\newcommand{\mytabref}[1]{table~\ref{#1}}
\newcommand{\myfigrefatstartofsentence}[1]{Figure~\ref{#1}}
\newcommand{\myfigref}[1]{figure~\ref{#1}}
\newcommand{\definmath}[2] {\def#1{\ifmmode#2\else$#2$\fi}}
\definmath\amin{\mathrm{min}}
\definmath{\cht}{{\tilde{\chi}}}
\definmath{\DeltaMChi}{{\Delta M_{\cht_1}}}
\def\mtx{{m_{TX}}}
\def\mtxsq{{m_{TX}^2}}
\definmath{\squark} {{\tilde{q}}}

\begin{document}

\vskip-15em\hskip-3em
  \underline{\small Cavendish HEP-2002-02/14}
\hskip58mm  \underline{\small PACS: 14.80.Ly 13.85.Qk}
\\\vskip5mm
{\Huge{$\mttwo$ : the truth behind the glamour}\vspace{7mm}}
\begin{center}\begin{tabular}{ccc}
\epsfig{file=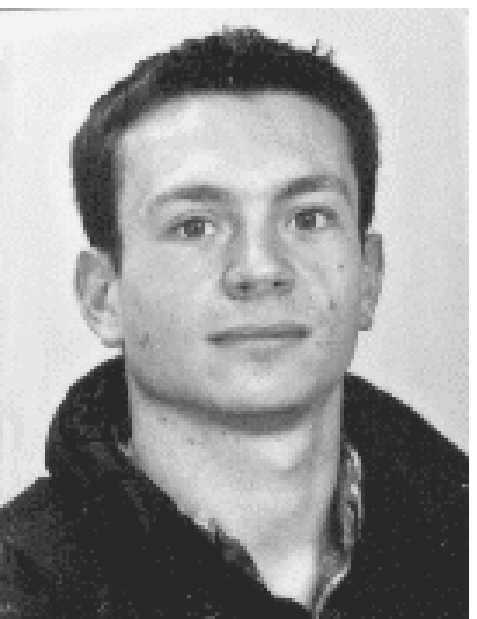, width=0.2\textwidth}&
\epsfig{file=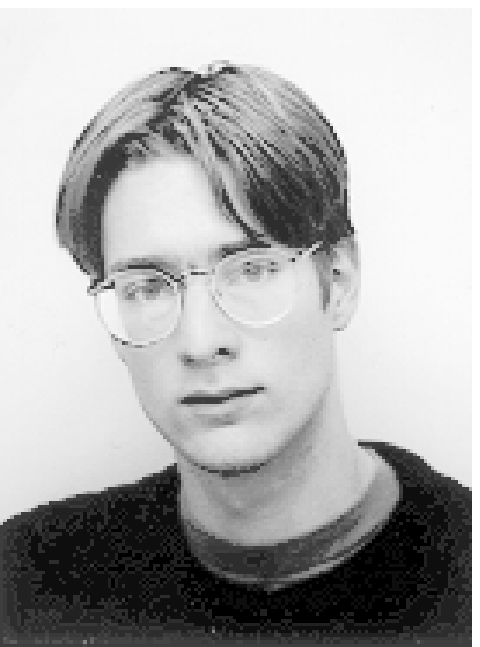, width=0.2\textwidth}&
\epsfig{file=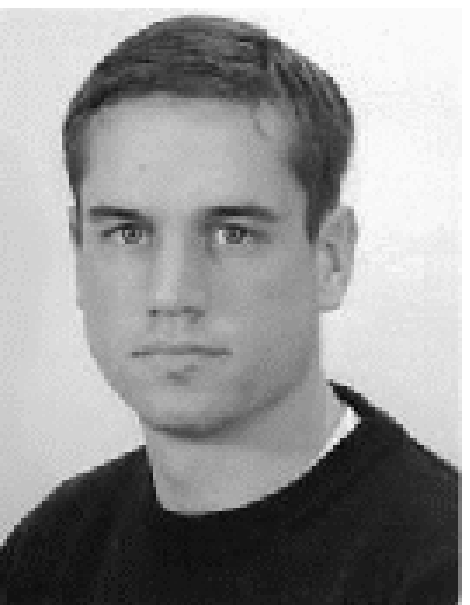, width=0.2\textwidth}\\
Alan Barr$^\dag$%\footnote{Email address: barr@hep.phy.cam.ac.uk} 
& Christopher Lester$^\ddag$ %\footnote{Email address: lester@hep.phy.cam.ac.uk}
& Phil Stephens$^\S$ %\footnote{Email address: stephens@hep.phy.cam.ac.uk}\\
\end{tabular}\\\vspace{3mm}
{\em Cavendish Laboratory, University of Cambridge, 
	Madingley Road, Cambridge,
        CB3\nolinebreak\ \nolinebreak{0HE,} UK}
\\\vspace{2mm}
Email address: $^\dag$barr@hep.phy.cam.ac.uk, 
$^\ddag$lester@hep.phy.cam.ac.uk, 
$^\S$stephens@hep.phy.cam.ac.uk
\end{center}
\vspace{2mm}
\begin{center}
\begin{minipage}{.9\linewidth}
{\bf
We present the kinematic variable, $\mttwo$, 
which is in some ways similar to the more familiar 
`transverse-mass', but which can be used in events
where two or more particles have escaped detection.
We define this variable and describe the event topologies
to which it applies, then present some of its mathematical properties.
We then briefly discuss two case studies which show how
$\mttwo$ is vital when reconstructing the masses of supersymmetric particles
in mSUGRA-like and AMSB-like scenarios at the Large Hadron Collider.
}
\end{minipage}
\end{center}
\section{Introduction}

\begin{figure}[t]\begin{center}
\epsfig{file=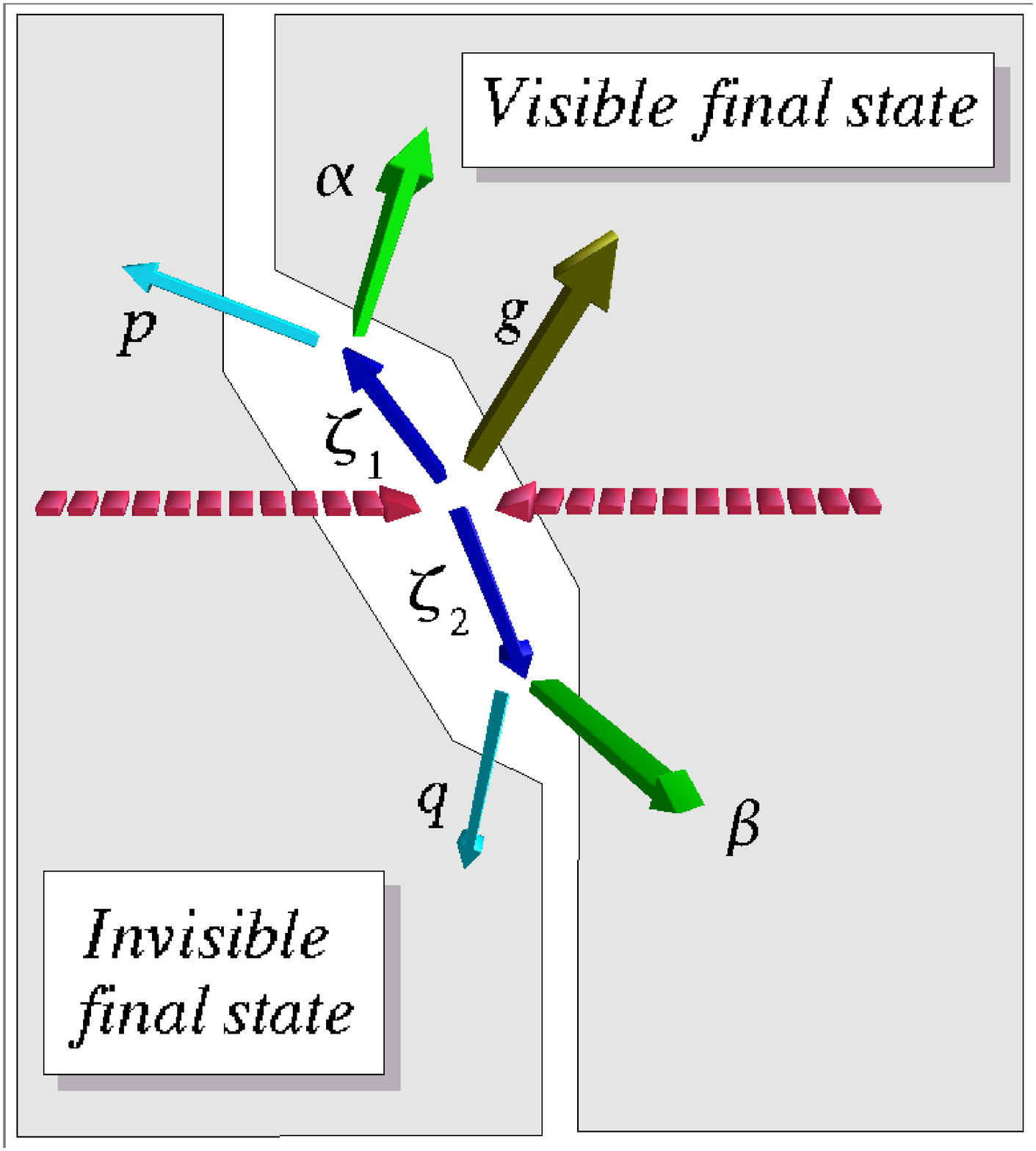,width=0.6\textwidth} \capbox{A simple pair
production event at the LHC}{Schematic representation of a simple
$R$-parity conserving event at the LHC in which \susic\ particles were
pair-produced.  The colliding protons are shown coming in from the
left and right.  The collision has pair produced two massive susy
particles, $\zeta_1$ and $\zeta_2$ (dark blue).  Each of these has
been shown decaying to something visible ($\alpha$ or $\beta$) and to
an undetected neutralino ($p$ or $q$).  The typical event will also
contain some initial- or final-state radiation, or other debris,
represented here by $g$.  In this figure it has been assumed that $g$
consists entirely of visible particles.  
%The diagram also displays
%some notation used in the text: $\Sigma$ is used to refer to the total
%visible transverse energy-momentum in the event, while $B$ denotes the
%total missing momentum transverse energy-momentum.  Because the center
%of mass energy of the collision ($\sqrt s$) is not known, the
%dependence of $B$ upon $\sqrt s$ must remain explicit in its
%definition.  The constant lorentz vector $\Lambda$, representing the
%transverse energy-momentum of a unit mass at rest in the lab frame,
%permits this dependence to be written down.  In the absence of sources
%of missing momentum other than $p$ and $q$ (the neutralinos) the
%assumtion is made that $B$ models possible values for
%$p+q$.
\label{fig:schematic}}
\end{center}\end{figure}

Reconstructing R-parity conserving supersymmetric events
will be difficult at the Large Hadronic Collider (LHC)
because of the following factors which limit our knowledge of the event:
\begin{itemize}
\item %{\bf (i)} 
two massive particles have escaped undetected,
\item %{\bf (ii)} 
the masses of these particles are unknown,
\item %{\bf (iii)} 
the masses of their `parent' particles are unknown,
\item %{\bf (iv)} 
the center-of-mass energy of the collision is not known, and
\item %{\bf (v)} 
the boost along the beam axis of the collision center-of-mass is not known either.
\end{itemize}

An example of such an event is shown schematically in Figure~\ref{fig:schematic}, 
where a pair of \susic\ particle have been produced, each of which has decayed to
some visible and some invisible daughters.

An important question to ask is ``What model-independent information
about sparticle masses can be deduced from events of this type?''.
The question can be seen to be a harder version of a number of older
problems with which we are more familiar.

Searches at the Large electron-positron collider (LEP) for pair produced sparticles 
have much in common with
this problem, although to first order they did not suffer from the
latter two of the above problems.  In some ways, the problem at the LHC is more
akin to that faced by the other hadronic collier experiments, 
such as UA1, UA2, CDF or DO where the $W$-mass has been measured from its 
decay to a lepton and a neutrino.  This has been achieved using the `transverse mass' event
variable, $m_T$, a variable which on an event-by-event basis generates
a lower bound for the $W$-mass, and the end-point of whose
distribution is the $W$-mass 
(\cite{Arnison:1983rp,Banner:1983jy,Affolder:2000bp,Abazov:2001wq}).
Their case was easier than that at the LHC,
however, as they only had {\bf one} such decay per event, and in
addition they could assume knowledge of the masses of both of the
particles into which the $W$ decayed.

\par

The approach to the problem first proposed in \cite{pubstransversemass} 
and subsequently developed in \cite{LESTER:THESIS} and \cite{Barr:2002ex} proposes the creation of a new kinematic variable,
$\mttwo$,\footnote{Because of its use in \susic\ events, $\mttwo$ has
acquired the nickname of the `stransverse' mass.}  analogous to the
transverse mass, whose kinematic endpoint carries model independent
information about (to first order) the mass difference between the
primary and the secondary supersymmetric particles.  Mention is made in
\cite{Barr:2002ex} of generalisations to this variable ($\mtthree$,
$\mtfour$, ...) which may be used when events contain extra missing
particles (\eg\ neutrinos) as well as the two neutralinos.

\par

The purpose of this article is not to discuss new physics results
which might be obtained with $\mttwo$ (for these the reader is encouraged to read
\cite{LESTER:THESIS} and \cite{Barr:2002ex}) but rather it aims to take a
closer look at more technical issues concerning the use and
interpretation of $\mttwo$, and its related variables.  It is hoped
that by concentrating information on $\mttwo$ in this way, this
article can act as a repository of $\mttwo$ `know-how' for future
investigations.

\section{A concrete example}

It is perhaps easiest to introduce and motivate the definition of the
Cambridge $\mttwo$ variable using a concrete example.  This allows the
ingredients that make up $\mttwo$ to be introduced, one at a time in
an almost `natural' way.  Readers who would prefer a `top down'
description of $\mttwo$, \ie\ a description which starts with a
definition and then works towards its consequences, are directed to
skip to \mysecref{sec:topDown} where this approach is taken.

\par

The concrete example which will be used here is taken from \cite{Barr:2002ex}.  
This paper considered an (anomaly mediated)
$R$-parity conserving \susic\ model whose key property was that it
predicted a lightest chargino nearly mass degenerate with the lightest
neutralino.  With particular choices of model parameters, the only
chargino decay mode available was:
\begin{equation}
\chginooneplus \to \ntlinoone \pi^+ \label{MTX:chgdecay}.
\end{equation}
Events containing two such decays, \ie\ events containing two
simultaneous decays of an unseen particle of unknown mass into another
invisible particle of unknown mass and visible particle, are exactly
the sort of events that we hope to analyse with $\mttwo$.  This we
shall now begin to do.

\par

Considering for the moment just one of the decays of the form
\myeqref{MTX:chgdecay}, one can write the Lorentz invariant statement
\begin{equation} m_\chginooneplus^2 = m_\pi^2 + m_\ntlinoone^2 + 
2 \Bigl[ E_T^\pi E_T^\ntlinoone \cosh(\Delta\eta)-{\bf
p}_T^\pi\cdot{\bf p}_T^\ntlinoone \Bigr]
\label{MT2:MCHIDEF}\end{equation}
where ${\bf p}_T^\pi$ and ${\bf p}_T^\ntlinoone$ indicate pion and
neutralino 2-vectors in the transverse plane, and the transverse
energies are defined by
\begin{equation}E_T^{\pi} = { \sqrt { ({\bf p}_T^{\pi}) {^2} + m_\pi^2 }}
\qquad \hbox{ and } \qquad {E_T^\ntlinoone} = { \sqrt{ ({\bf
p}_T^\ntlinoone){^2} + m_\ntlinoone^2 } }\ .
\label{MT2:ETDEF}\end{equation}
Also \begin{equation}\eta=\half
\log\biggl[\frac{E+p_z}{E-p_z}\biggr]\end{equation} is the true
rapidity, so that
\begin{equation}\tanh\eta=p_z/E\ ,\qquad \sinh\eta=p_z/E_T\ ,\qquad \cosh\eta=E/E_T .\end{equation}

\par

In a hadron collider, only the transverse components of a missing
particle's momentum can be inferred, so it is useful to define the
transverse mass,
\begin{equation}
m_T^2 ( {\bf p}^{\pi}_T, {\bf p}_T^\ntlinoone; m_\ntlinoone) \equiv { m_{\pi^+}^2 + m_\ntlinoone^2 +
2 ( E_T^{\pi} E_T^\ntlinoone - {\bf p}_T^{\pi}\cdot{\bf p}_T^\ntlinoone ) }
\label{MT2:MTDEF}\end{equation}
which, because $\cosh(x)\geq 1$, is less than or equal to the mass of
the lightest chargino, with equality only when the rapidity difference
between the neutralino and the pion, $\Delta\eta_{\ntlinoone\pi}$ is
zero.  All other $\Delta\eta$ lead to $m_T<m_\chginooneplus$, so {\em
if} we knew the neutralino momentum, we could use $m_T$ to give an
event by event lower bound on the lightest chargino mass.  $m_T$ was
has been used this way in the measurement of the $W^\pm$ mass.

\par

In events considered in this example, however, there are expected to
be {\em two} unseen lightest supersymmetric particles 
(LSPs).\footnote{Though there may also be other unseen particles -- see \mysecref{sec:MT2GEN}.}  
Since only the {\em sum} of
the missing transverse momentum of the two neutralinos is known, the
best that can be done is to evaluate the quantity
\begin{eqnarray}
{ \min_{\slashchar{{\bf q}}_T^{(1)} +
\slashchar{{\bf q}}_T^{(2)} = {\bf \pmiss}_T }} {\Bigl[ \max{ \Bigl\{
m_T^2({\bf p}_T^{\pi^{(1)}}, \slashchar{{\bf q}}_T^{(1)}; m_\ntlinoone) ,\
m_T^2({\bf p}_T^{\pi^{(2)}} , \slashchar{{\bf q}}_T^{(2)}; m_\ntlinoone) \Bigr\} }
\Bigr]} \qquad \label{MT2:MT2DEFalmost}
\end{eqnarray}
which is thus a {\em lower bound} on the square of the transverse mass,
$m_T$, for events where two decays of the type \myeqref{MTX:chgdecay}
occur.  Note that this minimisation has forced us to introduce a
pair of dummy two-vectors $\slashchar{{\bf q}}_T^{(1)}$ and
$\slashchar{{\bf q}}_T^{(2)}$ which, constrained by the minimisation
condition, parametrise our lack of knowledge about the two {\em true}
neutralino momenta.  Finally, we must recognise that under most
circumstances, the value of $m_\ntlinoone$ is unlikely to be known, or
may only be known with limited precision.  In order to make our
ignorance of $m_\ntlinoone$ explicit, we thus define a new free
parameter, $\chi$, calling it the `neutralino mass parameter',
intending it to denote any guess we might have as to the true
neutralino mass $m_\ntlinoone$.  Using it in place of $m_\chi$, we
convert \myeqref{MT2:MT2DEFalmost} into the following definition of a
new kinematic variable:
\begin{eqnarray}
\mttwosq(\chi) &\equiv& { \min_{\slashchar{{\bf q}}_T^{(1)} +
\slashchar{{\bf q}}_T^{(2)} = {\bf \pmiss}_T }} {\Bigl[ \max{ \Bigl\{
m_T^2({\bf p}_T^{\pi^{(1)}}, \slashchar{{\bf q}}_T^{(1)}; \chi) ,\
m_T^2({\bf p}_T^{\pi^{(2)}} , \slashchar{{\bf q}}_T^{(2)}; \chi) \Bigr\} }
\Bigr]}.\label{MT2:MT2DEF}
\end{eqnarray}

\par

The quantity defined in \myeqref{MT2:MT2DEF} is the Cambridge $\mttwo$
variable which is the subject of this
\label{sec:secwheremt2isfirstmotiv} document.

\par

Staying within the framework of this example, we can now go on to
describe some of the the desirable model-independent properties which
$\mttwo$ possesses.  

\subsection{Properties of $\mttwo(\chi)$.}

Firstly, is worth noting that the $\mttwo$ variable is not strictly a
`variable', and would more correctly be termed a `function', as it
retains a dependence on the unknown parameter $\chi$.  Ideally, $\chi$
would ideally be set equal to the mass of the missing heavy particle,
but in most of the situations in which the variable is likely to be
used, the mass of the invisible object is unlikely to be known, or may
only be known with a large uncertainty.  The $\chi$ dependence
remains, therefore.  A more detailed discussion of how this can
affects the use of $\mttwo$ takes place in
\mysecref{sec:forchidepenedence}.

\par

Secondly, {\em from its method of construction}, it is clear that for
any given event
\begin{eqnarray}
m_\pi+m_\ntlinoone & \le & \mttwo(m_\ntlinoone) \le m_\chginooneplus\label{eq:mtworange}, \qquad\rm{and} \\
m_\pi+\chi & \le & \mttwo(\chi)
\end{eqnarray}
It is certainly not immediately clear, however, that events can always
exits for which $\mttwo$ is capable of reaching all of these
endpoints.  In fact it turns out that such events do always exist, and
proof of this is given in
\mysecref{sec:secinwhichmt2rangeforidentdecaysisdetailed}.  So, having
defined the quantity $\mttwomax(\chi)$ by
\begin{eqnarray}
\mttwomax(\chi) & = &  \max_{\rm{many\ events}}\left[{\mttwo(\chi)}\right],
\end{eqnarray}
the important result to draw from all of this is that the upper
kinematic limit of $\mttwo$ satisfies
\begin{eqnarray}
\mttwomax(m_\ntlinoone) & = & m_\chginooneplus\label{eq:topoftherange}.
\end{eqnarray}
This is the main model-independent statement that $\mttwo$ is able to offer.

\subsection{Going beyond pairs of two body decays}

The scenario in which $\mttwo$ has been introduced, thus far, is
relatively simple; each event contains a pair of charginos, and each
of these decays via a two body decay into a charged pion and an unseen
neutralino. 
We will now consider in more detail what happens when:
\begin{itemize}
\item
the neutralinos are not the only missing particles, 
\item
the initial (\eg\ chargino) decays are not both two body decays, and
\item
$\mttwo(\chi)$ is evaluated at values of $\chi\ne m_\ntlinoone$.
\end{itemize}

\subsubsection{Extra missing particles and multi-particle decays}

\label{sec:MT2GEN}

\begin{figure}[t]
\begin{center}
\epsfig{file=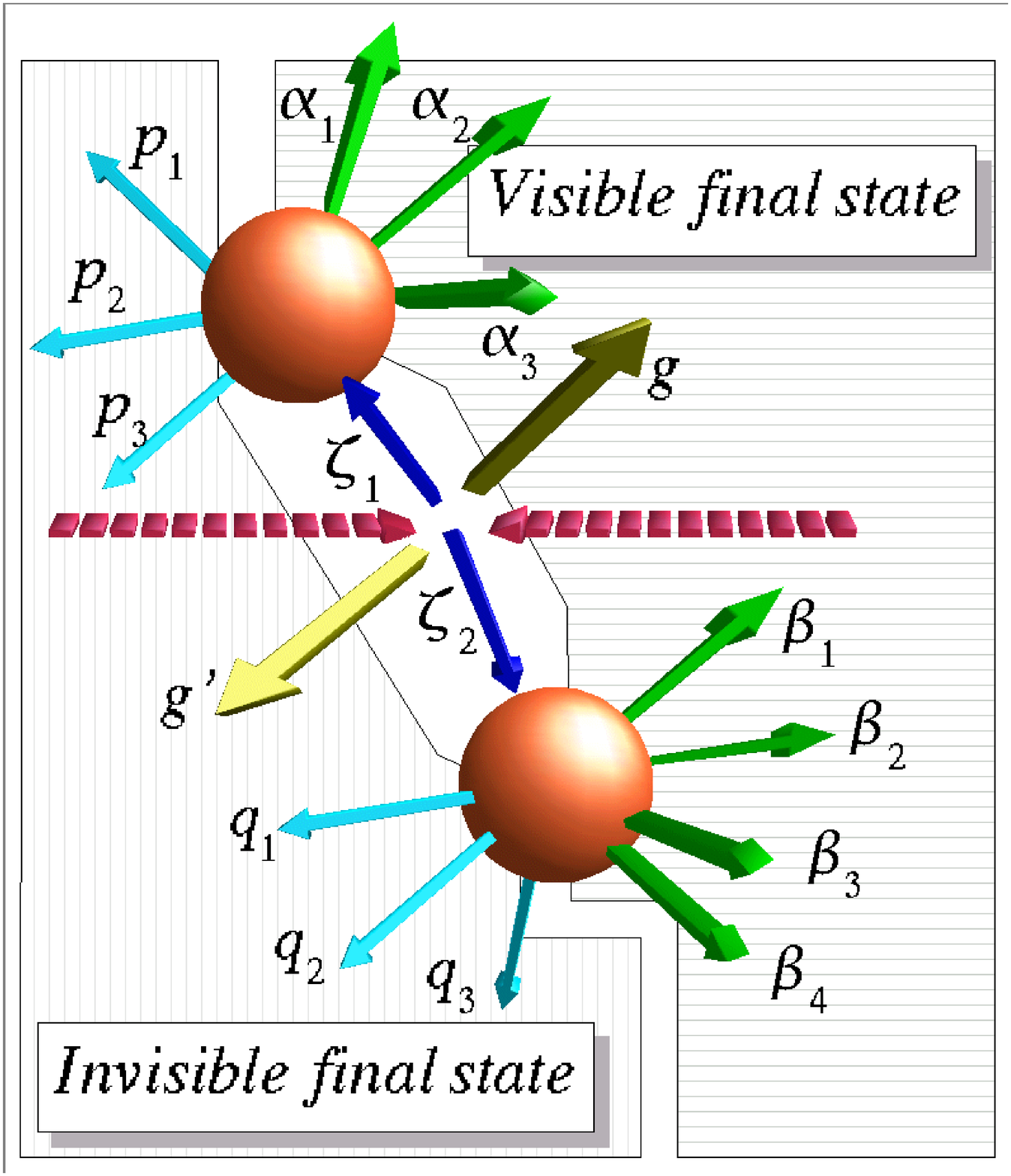,width=0.5\textwidth} \capbox{A typical pair
production event at the LHC}{Schematic representation of a
$R$-parity conserving event at the LHC in which \susic\ particles are
pair-produced.  The colliding protons are shown coming in from the
left and right.  The collision has pair produced two massive susy
particles, $\zeta_1$ and $\zeta_2$ (dark blue).  Each of these has
been shown decaying to a collection of visible particles ($\alpha_i$
or $\beta_i$) and to a set of undetected particles ($p_i$ or $q_i$).
The purpose of the large spherical blobs is to hide the details of the
decay process(es) involved; in principle they may contain anything,
from one large $n$-body decay, to $n-1$ successive two-body decays.
The typical event will also contain some initial- or final-state
radiation, or other debris, represented here by $g$ (the visible
component) and $g'$ (the invisible component).  Comments in the text
apply only principally to events in which $g'$ is small enough to be
neglected.\label{fig:mtxSchematic}}
\end{center}
\end{figure}

The need for $\mttwo$ to be adaptable to situations in which the
neutralinos are not the only unobserved final-state particles may
again be demonstrated using as an example the model of \cite{Barr:2002ex}.  In this model, there were found to be some regions
of parameter space in which three-body chargino decay,
\begin{equation}
\chginoonepm \to  l^\pm \nu \ntlinoone\label{MTX:threebodchgdecay},
\end{equation}
had a rate comparable to that of the two-body decay
\myeqref{MTX:chgdecay} which we have already seen.  Here, the presence
of the neutrino (or antineutrino) in the final state means that we
have even less information about the event than before.  Nevertheless,
one would like to benefit, if possible, from events in which one or
two of these decays occur in place of the usual two-body decays.  This
type of event is just one of the general class of events depicted in
\myfigref{fig:mtxSchematic}.

\par

\par

It is clear that one can immediately generalise the $\mttwo$ of
\myeqref{MT2:MT2DEF} to suit events like those in
\myfigref{fig:mtxSchematic} in the following way.  Define the new
variable $\mtx$ by:
\begin{eqnarray}
\mtxsq = \min_{\rm{consistent\ splittings}} \left[ 
\max{ \left\{ {
(\sum_i{\alpha_i}+\sum_j{p_j})^2
,  
(\sum_i{\beta_i}+\sum_j{q_j})^2
} \right\} }
\right].\label{eq:mtxisdefinedhere}
\end{eqnarray}
The phrase ``consistent splittings'', describing the constraint on the
overall minimisation, needs a little explanation.  There are two sets of
unknown momenta.  The first of these is ${\mathcal
F}=\{p_i\}\cup\{q_j\}$, containing the unknown momenta of all the
unobserved {\em final}-state particles.  The other set, ${\mathcal
H}=\{h_i\}$, contains the momenta of any on-mass-shell particles which
were present at an intermediate stage during the decays of the initial
pair of sparticles ($\zeta_1$ or $\zeta_2$) to their final
states.  In other words, $\mathcal H$ contains the momenta of any
intermediate particles {\em hidden} within the large blobs in
\myfigref{fig:mtxSchematic}.  Minimisation over ``consistent
splittings'', then, means minimisation over all $p_i,q_j\in{\mathcal
F}$ and all $h_i\in{\mathcal H}$ subject to:
\begin{itemize}
\item
all $p_i$, $q_i$ and $h_i$ being on their respective mass-shells,
\item
momenta being conserved at all `hidden' vertices in which a short lived
intermediate particle with momentum $h_i\in{\mathcal H}$ decays, and
\item
the transverse components of $B=\sum_i{p_i}+\sum_j{q_j}$ being
consistent with the measured missing momentum $\Ptmiss$.
\end{itemize}
It is because the last of these requirements that we need events in
which $g'$, the momentum carried by any invisible particles which are not
descendants of a \susic\ particle, (see \myfigref{fig:mtxSchematic}) is negligible.  Were
there to be a large tail in the distribution of $g'$, this would
degrade the performance of $\mtx$ and $\mttwo$.

\subsubsection*{Example}

\begin{figure}[t]
\begin{center}
\epsfig{file=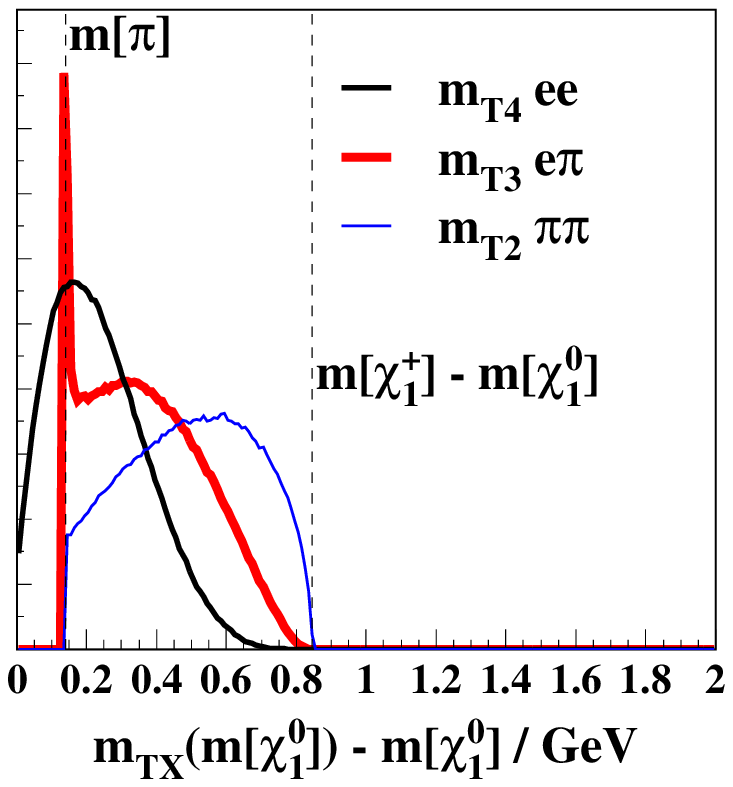,width=0.5\textwidth} \capbox{Example
distributions of $\mtx-m_\ntlinoone$.}{Simulations of
$\mtx(m_\ntlinoone)-m_\ntlinoone$ for $X=2,3,4$ using a simple phase-space
Monte-Carlo generator program for a pair of decays
$\squark\to\chginooneplus q$ followed by
$\chginooneplus\to\ntlinoone~\pi$ or
$\chginooneplus\to\ntlinoone~e~\nu_e$.  As the number of invisible
particles increases, the proportion of events near the upper limit
decreases.  Within the figure, subscripts are indicated by square brackets.\label{fig:deltamtxhisto}}
\end{center}
\end{figure}

We illustrate the remarks of the previous section by returning to the
example of \cite{Barr:2002ex} in the case where where charginos could
decay either by the two-body decay of \myeqref{MTX:chgdecay} or the
three-body decay \myeqref{MTX:threebodchgdecay}.  We can categorise
events in this scenario by the number of missing particles in the
event.  When both charginos decay via \myeqref{MTX:chgdecay} we only
have {\em two} missing particles (the neutralinos).  For each
three-body decay which takes the place of one of these two-body decays
we gain an {\em extra} missing particle in the form of a neutrino (or
antineutrino).  In short, the three categories of events could be
summarised as those containing one of the following: 
\[\chginoonepm
\chginoonepm\rightarrow\{ \pi^\pm \ntlinoone \pi^\pm \ntlinoone,\ {\rm
or\ } e^\pm \nu \ntlinoone \pi^\pm \ntlinoone,\ {\rm or\ } e^\pm \nu
\ntlinoone e^\pm \nu \ntlinoone \}\ . \]
The events had been produced by a
phase-space-only Monte-Carlo generator.  Three distributions of the quantity
$\mtx$, defined in \myeqref{eq:mtxisdefinedhere}, were then generated
from each of these sets of events.  Using the number of missing
particles to categorise these events, the values of $\mtx$ measured in
each case are referred to as $\mttwo$, $\mtthree$ and $\mtfour$. 
The resulting distributions for $\mtx(m_\ntlinoone)-m_\ntlinoone$ are shown in
\myfigref{fig:deltamtxhisto}.

\par

It has already been mentioned that a key property of $\mttwo$ is that
the kinematic endpoint of its distribution occurs at
$\mttwomax(m_\ntlinoone)=m_\chginooneplus$ and so it is reassuring to
see in \myfigref{fig:deltamtxhisto} that a large number of events
reach this endpoint.  In the vicinity of the endpoint, the edge is seen
to be sharp and near vertical.  This shows that at the partonic level
a measurement of $\mttwomax$ would provide an excellent constraint on
the masses of the sparticles involved.  In section \ref{sec:realmt2} plots
from \cite{LESTER:THESIS, Barr:2002ex}, which include realistic
detector effects, will show that the subsequent smearing of the
$\mttwo$ edge, while significant, is still small.

\par

Looking next at the $\mtthree$ and $\mtfour$ distributions, it is
clear that the event fall-off in the vicinity of their kinematic
endpoints is much less steep than in the case of $\mttwo$.  This is
hardly surprising, given the reduced amount of information available
in these events.  Later, in
\mysecref{sec:numbersofeventsnearthetails}, the relative fraction of
events in the vicinity of the edge will be seen, more quantitatively,
to be due to the larger number of simultaneous conditions that events
near the edge must satisfy.  Although the endpoint, itself, becomes
increasingly harder to detect as the number of missing particles
increases, the $\mtx$ distributions are {\em all} capable of inferring
the mass scale associated with (in this case)
$m_\chginooneplus=m_\ntlinoone$ from the overall widths of their
distributions, which each scale with the endpoint position, albeit with
some dependence on the decays themselves, and on factors such as the
detector acceptance over the width of the distribution.

\par

Finally, one notes that the $\mtthree$ distribution has sharp peak at
$\mtthree=m_\ntlinoone+m_\pi$, not seen in the $\mttwo$ and $\mtfour$
distributions.  It will be shown later, in \mysecref{sec:dfgdkferhg}, that
this is an effect which can occur whenever the hypothesised decays on
each side of the event are different.\footnote{`Different', in this
context, means `being such that the minimum total invariant mass
attainable by the particles on one side of the event is not equal to
the minimum total invariant mass attainable by the particles on the
other side of the event'.  This happens principally when the particle
content of each decay differs.  In the case of $\mtthree$ in,
the AMSB example scenario, the two dissimilar minima are $ m_\pi +
m_\ntlinoone$ and $m_e + m_\nu + m_\ntlinoone$.}

\subsubsection{Other values of $\chi$}
\label{sec:forchidepenedence}
Now we return to a brief look at the effect of evaluating $\mtx$
distributions at values of $\chi$ different to the true neutralino
mass.  

\par

\begin{figure}[t]\begin{center}
\epsfig{file=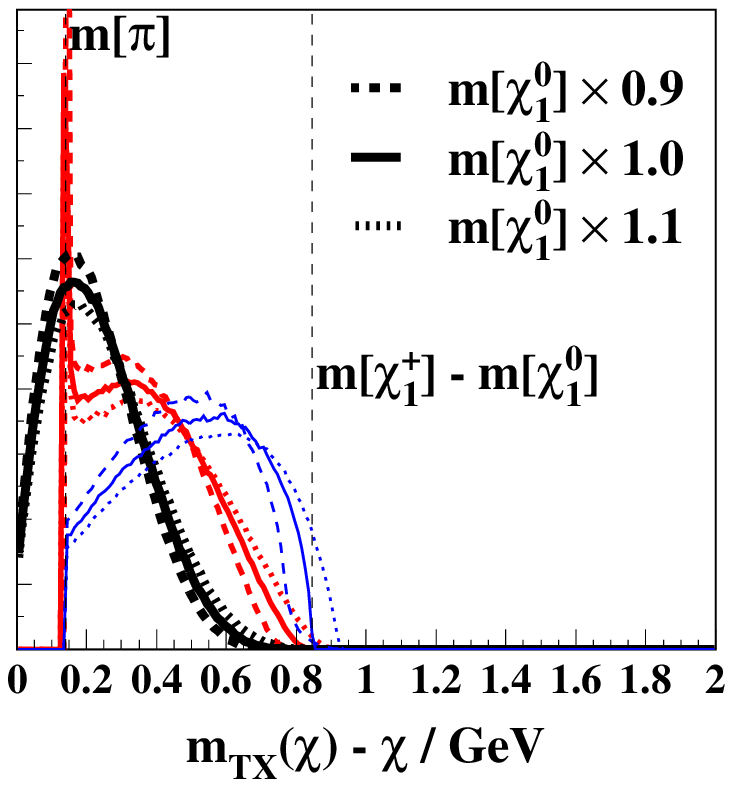,width=0.4\textwidth} \capbox{Variation of
$\mtx$ distributions with $\chi$.}{The distortion of $\mtx(\chi)-\chi$
when the LSP mass parameter, $\chi$, is varied by $\pm$~10\% about the
`ideal' value of $m_\ntlinoone$.  These curves show that
$\mtx(\chi)-\chi$ remains sensitive to the mass difference
$\DeltaMChi=m_\chginooneplus-m_\ntlinoone$.  In this simulation
$\DeltaMChi=0.845$~GeV, $m_\ntlinoone=161.6$~GeV, and the electron and
neutrino mass were neglected.  The normalisation is arbitrary.  Within
the figure, subscripts are indicated by square brackets.
\label{MTX:MTXVERSUSCHIPLOTS}
}
\end{center}
\end{figure}

\begin{figure}[t]\begin{center}
\epsfig{file=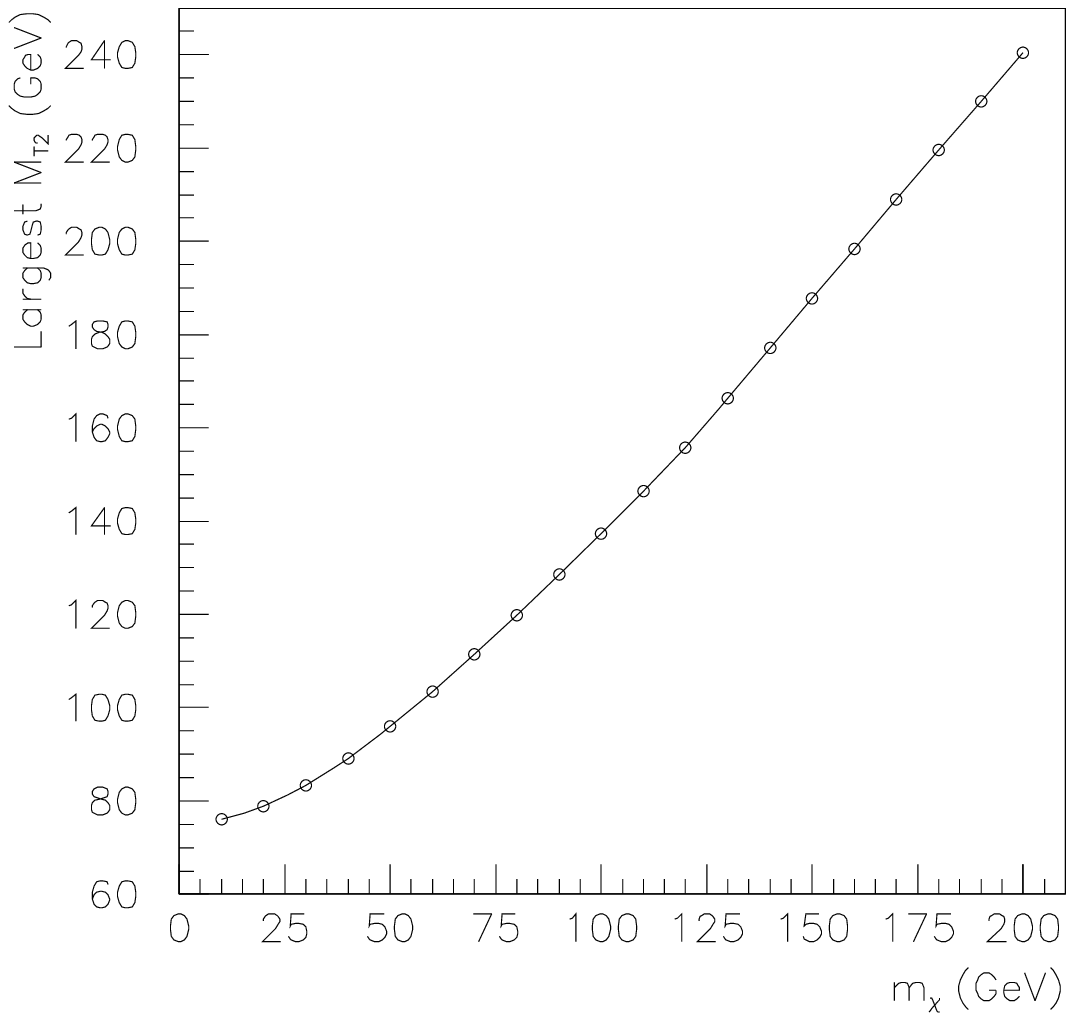,width=0.6\textwidth} \capbox{Variation
of $\mttwomax(\chi)$ with $\chi$}{Variation
of $\mttwomax(\chi)$ with $\chi$ for a set of $\tilde l^+\tilde
l^-\rightarrow l^+\ntlinoone l^-\ntlinoone$ events generated by a phase-space
Monte-Carlo using $m_{\tilde{l}}=157.1$~GeV and
$m_{\ntlinoone}=121.5$~GeV.  Note that $\mttwomax(\chi)-\chi$
decreases as $\chi$ increases.  The asymptote has unit
gradient.\label{fig:mlvsmchi} }
\end{center}
\end{figure}

\par

\myfigrefatstartofsentence{MTX:MTXVERSUSCHIPLOTS} shows the same data
as in \myfigref{fig:deltamtxhisto}, but in addition it shows the
distributions that would be obtained by evaluating $\mtx(\chi)$ at
values of $\chi=m_\ntlinoone\pm10\%$.  In this particular example, 
where $m_\ntlinoone=162$~GeV, 
10\% (16~GeV) errors in $\chi$ result in similar {\em fractional}
errors in \DeltaMChi\ \ie\ of a few tens of MeV.
This shows that $\mttwo$ can be sensitive to small mass differences.
In this example, too, we see a {\em positive} correlation between the change in $\chi$
and the change in the position of the endpoint.  These examples are
not always typical, however.  For example, in
\cite{pubstransversemass} the authors considered $\mttwo$ in the
context of a double slepton decay to lepton a neutralino at SUGRA
Point 5, one of the five supergravity points proposed at
\cite{sugrapts} and described in \cite{sugrapt5}.  In this model, the
difference in mass between the decaying and final sparticles
($157.1-121.5=35.6$~GeV) is approximately $40$ times larger than in
the AMSB case, and so at SUGRA Point 5 we see a {\em negative}
correlation between changes in $\chi$ and $\mttwo(\chi)-\chi$. This is
illustrated in \myfigref{fig:mlvsmchi}.  Differing kinds of behaviour,
such as these, are typical of a variable like $\mttwo$ which has input
scales (\eg\ $m_\pi$, $m_\chginooneplus$, $\chi$ and
$m_\chginooneplus-\chi$ ) which can have a large number of relative
hierarchies associated with them.  For example, the AMSB scenario has
$m_\pi\approx m_\chginooneplus-\chi \ll \chi \leq m_\chginooneplus$,
while SUGRA Point 5 has $m_l \ll m_{\tilde{l}}-\chi \approx \chi <
m_{\tilde{l}}$.

\par

\begin{figure}[t]
\begin{center}
\epsfig{file=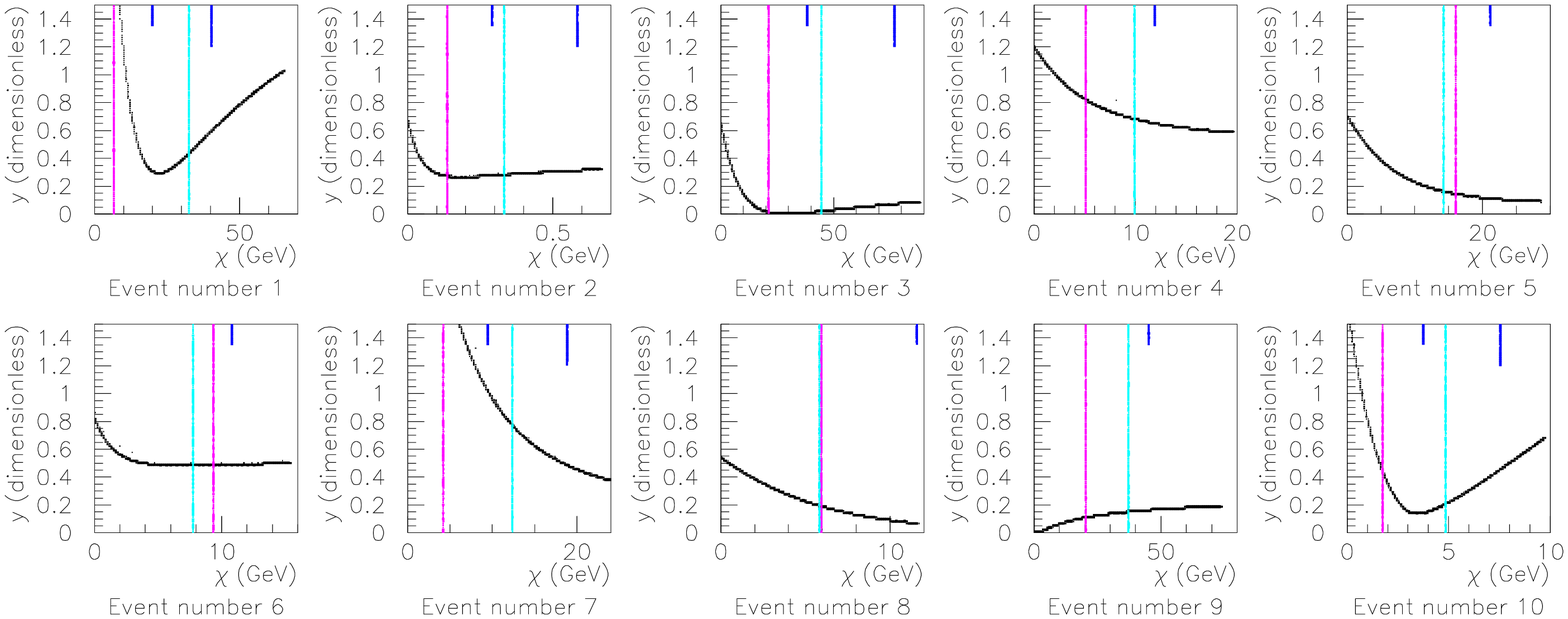,width=0.99\textwidth}
\capbox{$\chi$ dependence of $\mttwo$ for random events}{These plots
show examples of how $y(\chi)$, defined in \myeqref{eq:defOfythefrac},
can depend on $\chi$.  The plots were generated using the following
procedure.  Ten sets of masses satisfying $m_\pi+m_\ntlinoone <
m_\chginooneplus$ were randomly generated.  According to each set of
masses, a phase-space Monte-Carlo generated a single event of the type shown in
\myfigref{fig:schematic} containing two $\chginoonepm\rightarrow
\pi^\pm \ntlinoone$ decays.  The plots above show how, in each event,
the value of $y(\chi)$ (a dimensionless rescaling of $\mttwo(\chi)$)
depended upon $\chi$ over the range $0<\chi<2 m_\ntlinoone$.  The true
value of the neutralino mass (\ie\ that used in the Monte-Carlo for the
decay) is marked by the vertical line at the center of each plot
(cyan), while the other vertical line marks the value of the pion mass
(magenta).  The short and long vertical ticks (dark blue) mark
$m_\chginooneplus/2$ and $m_\chginooneplus$ respectively.
\label{fig:prettyChiVariation} }
\end{center}
\end{figure}

\par

We now take a final look at how $\mttwo$ depends upon $\chi$ by
looking not just at events near the kinematic endpoint, but at events
in general.  To help, we define $y(\chi)$, a rescaling of
$\mttwo(\chi)$, as follows:
\begin{equation}
y(\chi)\equiv
\frac{\mttwo(\chi)-\chi-m_\pi}{m_\chginooneplus-m_\ntlinoone-m_\pi}.\label{eq:defOfythefrac}
\end{equation}
By looking back at \myeqref{eq:mtworange} and
\myeqref{eq:topoftherange}, this variable can be seen to map
$\mttwo(m_\ntlinoone)$ into the range $[0, 1]$.\footnote{Note that
$\mttwo(\chi)$ is in general {\bf not} constrained to lie in this
range.}  This makes it easier to compare values of $\mttwo$ coming
from events with different sparticle masses.  A value of
$y(m_\ntlinoone)$ close to 0 (or 1) indicates an event close to the
lower (or upper) kinematic endpoint of the $\mttwo(m_\ntlinoone)$
distribution.  \myfigrefatstartofsentence{fig:prettyChiVariation}
shows how $y(\chi)$ varies with $\chi$ for ten random events, each
generated using a random set of masses satisfying $m_\pi+m_\ntlinoone
< m_\chginooneplus$ as described in the figure caption.  The main
conclusion to draw from these plots is that there is no easy way to
say, in advance, how $\mttwo(\chi)$ will vary with $\chi$ in a given
event, even in the vicinity of $m_\ntlinoone$.  In general
$\mttwo(\chi)$ can rise, fall, or even be stationary w.r.t to $\chi$
near $m_\ntlinoone$, depending on the masses of the particles involved
in the decays.

\section{Some mathematical results concerning the variable $\mtx$}
\label{sec:topDown}

%When considering events such as these at the LHC, it has been found
%{\bf CITE ORIGINAL MT2 PAPER} to be useful to construct a variable
%analagous to the transverse mass as follows:

In \mysecref{sec:secwheremt2isfirstmotiv}, $\mttwo$ and its friends
were defined using lab-frame momenta and with minimisation
conditions (such as $\slashchar{{\bf q}}_T^{(1)} + \slashchar{{\bf
q}}_T^{(2)} = {\bf \pmiss}_T $) not specified in lorentz-invariant
forms.  The definition of $\mttwo$ from  \mysecref{sec:secwheremt2isfirstmotiv} may be summarised as follows:
\begin{eqnarray}
{ \mttwosq (\PtlepOne, \PtlepTwo, \Ptmiss; \chi)} \equiv {
\min_{\slptwo_1 +\slptwo_2 = \Ptmiss }{ \Bigl[ \max{ \{
m_T^2(\PtlepOne, \slptwo_1; \chi) , m_T^2(\PtlepTwo,
\slptwo_2; \chi ) \} } \Bigr] } },\label{eq:mt2def1}
\end{eqnarray}
where
\begin{eqnarray}
m_T^2(\ptwo_T^l,\ptwo_T^{\tilde{\chi}};\chi) \equiv m_l^2 +
\chi^2 + 2 ( E_T^l E_T^{\tilde{\chi}} -
\ptwo_T^l\cdot\ptwo_T^{\tilde{\chi}}),
\label{eq:mtd}
\end{eqnarray}
in which $E_T^l=\sqrt{{\left|{{\bf p}_T^l}\right|}^2 + m_l^2}$,
$E_T^{\tilde\chi}=\sqrt{ {\left|{{\bf p}_T^{\tilde\chi}}\right|}^2 +
\chi^2}$.  In the following sections we will replace these definitions
by equivalent, but explicitly lorentz-invariant ones, which are easier
to manipulate mathematically.

\subsection{Definitions}

The natural way to write $\mttwo$ in a manifestly $(1,2)$-lorentz invariant form is as follows:
\begin{eqnarray}
{ \mttwosq (\alpha, \beta, \Sigma, \Lambda; \chi)} \equiv {
\min_{ \left\{ {\smallTwoVec{\phantom{p}p+q=\sqrt s
\Lambda-\Sigma\phantom{A}}{p^2=q^2=\chi^2} } \right\} } { \Bigl[ \max{
\{ (\alpha + p)^2 , (\beta + q)^2 \} } \Bigr] } }.\label{eq:mt2def2}
\end{eqnarray}

Here, $\mttwo$ has been written as a function of the four
$1+2$-dimensional lorentz vectors which describe each event ($\alpha$, $\beta$,
$\Sigma$ and $\Lambda$) and one real parameter $\chi$.
The transverse
lorentz vectors of the two visible particles coming from each of the
hidden decays are represented by $\alpha$ and $\beta$, while $\Sigma$
represents the total transverse energy-momentum seen in the event.
This is consistent with the notation used in
Figure~\ref{fig:schematic}.  The only new vector, $\Lambda$, defines
the laboratory frame by being the $(1,2)$-energy-momentum of a
particle of unit mass ($\Lambda^2=1$) at rest in the laboratory.  The
total transverse momentum of the event (visible and invisible) can
only be assumed to be zero in the laboratory frame, and so knowledge
of how to boost to the laboratory frame is essential. This is why
$\Lambda$ is needed.

\par

Note that (\ref{eq:mt2def2}) includes a minimisation over $\sqrt s$, a
parameter which accounts for our lack of knowledge of the
center-of-mass energy of the whole event.  The requirement that the
hypothesised neutralino momenta are real, \ie\ $(\sqrt s
\Lambda-\Sigma)^2\ge (2 \chi)^2$, constrains $\sqrt s$ to be
chosen from the region in which 
\begin{eqnarray}
\sqrt s\ge \Lambda . \Sigma + \sqrt{4
\chi^2 + ((\Lambda . \Sigma)^2-\Sigma^2)}.\label{eq:rootsrange}
\end{eqnarray}

\par

Similarly, one can also define $\mtx$ in a manifestly lorenz-invariant
form:
\begin{eqnarray}
{ \mtxsq (\hat\alpha, \hat\beta, \Sigma, \Lambda; \chi)} \equiv \min
{ \Bigl[ \max{ \{ (\hat\alpha + \hat p)^2 , (\hat\beta + \hat q)^2 \}
} \Bigr] } \qquad\rm{over}\nonumber\\ { \left\{ {
\smallThreeVec{\phantom{p}\hat p+\hat q=\sqrt s
\Lambda-\Sigma,\phantom{A}}{\phantom{A}{p_i, q_i\ }{\rm{and\ }}{ h_i\ }\rm{all\ on\
their\ mass\ shells,\ and\phantom{A}}} {{\phantom{p}\rm{momenta\ conserved\ at\ all\ internal}}\ h\
\rm{decays}\phantom{A}} } \right\} }\label{eq:mtxdefboguydf},
\end{eqnarray}
in which the same notation has been used as in
\myfigref{fig:mtxSchematic} and equation \myeqref{eq:mtxisdefinedhere}, and
in which the `hats' indicate summation over all vectors of a set (\eg\
$\hat \alpha = \sum_i \alpha_i$).

\par

If desired, one may remove the `$\max$' at the expense of moving to
lorentz {\em four}-vectors and remembering to minimise over all possible longitudinal
boost of the center of momentum (here denoted by the lotentx boost $L_z$):
\begin{eqnarray}
{ \mttwosq (\alpha, \beta, \Sigma, \Lambda; \chi)} &\equiv& {
\min_{ \left\{ {\smallThreeVec{D^2=D_1^2=D_2^2}{\phantom{p}p+q=\sqrt s
L_z \Lambda-\Sigma\phantom{A}}{p^2=q^2=\chi^2} } \right\} } 
{ \Bigl[ 
D^2
\Bigr] } },\qquad{\rm where}\label{eq:mt2def2different}\\
D_1^2&=&(\alpha + p)^2,\qquad{\rm and}\\
D_2^2&=&(\beta + q)^2.
\end{eqnarray}
This way of representing $\mttwo$ most clearly captures the spirit
in which it provides an event-by-event lower bound on the
initial sparticle mass.

\subsection{Results concerning $\mttwo$}

%In terms of transverse momenta measured in the laboratory frame,
%(\ref{eq:mt2def2}) is equivalent to:
%
%\par

In the case where both visible particles have the same mass, \ie\ in
the case where $\alpha^2=\beta^2=m_l^2$, (\ref{eq:mt2def2}) may be
re-written in the form:
\begin{eqnarray}
{ \mttwosq'(\alpha, \beta, \Sigma, \Lambda; \chi)} \equiv
 m_l^2 + \chi^2+{ \min_{ \left\{ {\smallTwoVec{\phantom{p}p+q=\sqrt s
 \Lambda-\Sigma\phantom{A}}{p^2=q^2=\chi^2} } \right\} } { \Bigl[ 2 \max{
 \{ \alpha . p , \beta . q \} } \Bigr] } }.\label{eq:mt2def3}
\end{eqnarray}

\par

It was shown in \cite{Barr:2002ex} that the solution of
(\ref{eq:mt2def3}) must select vectors $p$ and $q$ for which $\alpha
. p = \beta . q$.  Using this information, one may perform half of the
minimisation in (\ref{eq:mt2def3}) analytically.  This allows
(\ref{eq:mt2def3}) to be re-written as a minimisation over a single
real variable, $\sqrt s$, as follows:
\begin{eqnarray}
{ \mttwosq'(\alpha, \beta, \Sigma, \Lambda; \chi)} \equiv
{m_l^2 + \chi^2+\half \min_{\sqrt s} {\left[ {(\sigma . B) \mybigq - \sqrt{\sigma^2 \mybigq - 4 m_l^2}\sqrt{B^2 \mybigq-4 \chi^2}} \right]}},\phantom{Ma}\label{eq:mt2def4}
\end{eqnarray}
where
\begin{eqnarray}
\mybigq  &=&  1-\frac{(\mynabla . B)^2}{(\sigma . B)^2-\sigma^2 B^2}, \qquad(\Rightarrow 0\le\mybigq\le1)\\
\sigma &=& \alpha+\beta,\\
\mynabla &=& \alpha-\beta, \qquad \rm{and}\\
B &=&  \sqrt s \Lambda - \Sigma. \label{eq:bIsDefined}
\end{eqnarray}
We should note that the constraint which has just been imposed, namely
$\alpha . p = \beta . q$, is more stringent than the $\sqrt s$
constraint (\ref{eq:rootsrange}) which was only there to ensure that
the hypothesised neutralinos were not tachyonic.  As a
consequence, the range over which $\sqrt s$ may be varied when
performing the minimisation in (\ref{eq:mt2def4}) must be replaced by
the stronger condition that each of the quantities under radicals in
(\ref{eq:mt2def4}) be positive.

\par

It is interesting to note that if we define two new transverse lorentz vectors ($\overline\sigma$ and $\overline B$) via a rescaling of existing transverse lorentz vectors according to
\begin{eqnarray}
\overline\sigma &=& \sigma \sqrt\mybigq ,  \qquad \rm{and}\\
\overline B &=& B\sqrt\mybigq, 
\end{eqnarray}
then we can rewrite
(\ref{eq:mt2def4}) in the form
\begin{eqnarray}
{ \mttwosq'(\alpha, \beta, \Sigma, \Lambda; \chi)} \equiv
{m_l^2 + \chi^2+\half \min_{\sqrt s} {\left[ {(\overline \sigma . \overline B)  - \sqrt{{\overline\sigma}^2  - 4 m_l^2}\sqrt{{\overline B}^2 -4 \chi^2}} \right]}}.\label{eq:mt2def4aa}
\end{eqnarray}
This is not much of an improvement in itself, but it motivates the
definition of two new lorentz {\bf four}-vectors;
\begin{eqnarray}
\tilde\sigma &=& (\overline \sigma, \sqrt{{\overline\sigma}^2  - 4 m_l^2}),  \qquad \rm{and}\\
\tilde B &=& (\overline B, \sqrt{{\overline B}^2  - 4 \chi^2}),
\end{eqnarray}
which we see, by construction, satisfy the following fixed-mass relations:
\begin{eqnarray}
m_{\tilde \sigma} &=& \sqrt{{\tilde \sigma} ^ 2}  =  2 m_l\label{eq:constmassrel1},  \qquad \rm{and}\\
m_{\tilde B} &=& \sqrt{{\tilde B}^2}  =  2 \chi\label{eq:constmassrel2}.
\end{eqnarray}
In terms of these new lorentz four-vectors, then, we can finally re-write 
(\ref{eq:mt2def4aa}) as 
\begin{eqnarray}
{ \mttwosq'(\alpha, \beta, \Sigma, \Lambda; \chi)} &\equiv&
{m_l^2 + \chi^2+\half \min_{\sqrt s}\left({\tilde\sigma . \tilde B}\right)}, \qquad\rm{or}\label{eq:mt2def4bb}\\
{ \mttwo'(\alpha, \beta, \Sigma, \Lambda; \chi)} &\equiv&
{\half \min_{\sqrt s}\left|{\tilde\sigma+\tilde B}\right|}\label{eq:mt2def4cc}.
\end{eqnarray}
It is interesting to note that the constant mass relations
(\ref{eq:constmassrel1}) and (\ref{eq:constmassrel2}), taken together
with the definition of $\mttwo'$ shown in (\ref{eq:mt2def4cc}), make
it self evident that the value of $\mttwo'$ obtained in a given event
is bounded below by $m_l+\chi$, as expected.

\subsubsection*{Approximations}

To get a better idea of the way in which $\mttwo$ depends on its
inputs, one might hope to find a concise closed-form analytic
definition of the variable.  Thus far, however, $\mttwo$ and $\mttwo'$
have resisted all attempts to write them in forms simpler than
(\ref{eq:mt2def1}), (\ref{eq:mt2def2}) and (\ref{eq:mt2def4}), except
in a few special cases.
For example, in the special case of events in which
the spatial part of the total visible transverse momentum is seen to
be zero in the laboratory frame (\ie\ events for which $(\Sigma
. \Lambda)^2=\Sigma^2$) one can show that (\ref{eq:mt2def4}) is
equivalent to:
\begin{eqnarray}
{\mttwosq''(\alpha, \beta, \Lambda; \chi)} &\equiv& {m_l^2 +
\chi^2 + \chi\sqrt{4(\alpha . \Lambda)(\beta
. \Lambda) - (-\mynabla^2)}} \label{eq:noBoostNeeded} \\
(&=& \nonumber
{m_l^2 +\chi^2 + \chi\sqrt{2( E_\alpha E_\beta+m_l^2+\Ptalpha . \Ptbeta)}} \\
& & \mbox{in the laboratory frame}\ \ )\nonumber.
\end{eqnarray}

The limit of validity of \myeqref{eq:noBoostNeeded} can be explored as follows.
The laboratory frame energy that this special case solution assigns to $p$ and $q$ is given by
\begin{eqnarray}
p . \Lambda = q . \Lambda = \frac{(\sigma . \Lambda )\chi}{\sqrt{4(\alpha . \Lambda)(\beta . \Lambda)-(-\mynabla^2)}},
\end{eqnarray}
and so the velocity of the boost needed to take the laboratory frame
to the one in which the invisible particles are back to back could be
written, in this special case, as:
\begin{eqnarray}
(\ 0=\ )\qquad\left|{\bf v}\right|^2 &=& {\bf p}_\Sigma^2 / ( p.\Lambda + q.\Lambda )^2 \label{eq:vSquaredForNoBoost}\\
&=&
\frac
{
\left\{(\Sigma . \Lambda)^2-\Sigma^2\right\} 
\left\{4(\alpha . \Lambda)(\beta . \Lambda)-(-\mynabla^2)\right\}
}
{4 (\sigma . \Lambda )^2\chi^2}.\label{eq:vSquaredForNoBoostLorInv}\\
(&=& \frac{  {\bf p}_\Sigma^2 (E_\alpha E_\beta+m_l^2+\Ptalpha . \Ptbeta) }
{2 (E_\alpha+E_\beta)^2 \chi^2}\label{eq:vSquaredForNoBoostLast}\\
& & \mbox{in the laboratory frame}\ \ )\nonumber
\end{eqnarray}
In the light of the above, we can interpret (\ref{eq:noBoostNeeded})
as the leading term in an expansion of $\mttwosq'$ in powers of 
$\left|{\bf v}\right|^2$ as defined in
%(\ref{eq:vSquaredForNoBoost}), 
(\ref{eq:vSquaredForNoBoostLorInv}).
%or (\ref{eq:vSquaredForNoBoostLast}).
Given a particular event, all the quantities in
(\ref{eq:vSquaredForNoBoostLorInv}) may be evaluated, so one can
safely use (\ref{eq:noBoostNeeded}) to evaluate $\mttwosq'$ for events
in which $\left|{\bf v}\right|^2$ is observed to satisfy $\left|{\bf
v}\right|^2\ll 1$.\footnote{The reader is warned not to mistake
$\left|{\bf v}\right|$ for the speed associated with an actual boost
(real or conjectured) connected with the neutralino pair; $\left|{\bf
v}\right|$ could, for example, even exceed the speed of light if
$\chi$ were made sufficiently small!  It should only be assumed that
as $\left|{\bf v}\right|\rightarrow 0$, $\left|{\bf v}\right|$ will
tend to the speed, in the laboratory frame, associated with the
energy-momentum vector $B$.  (This is the $B$ which was originally
defined in (\ref{eq:bIsDefined}) and whose $\sqrt s$ value was
selected by the minimisation process in (\ref{eq:mt2def4}).)}

\subsection{Extremal values of $\mttwo$}

In this section, we show that the maximum value which
$\mttwo(m_\ntlinoone)$ can attain, for a given set of particle masses,
is indeed the mass of the initial sparticle.\footnote{Up to this
point, within the context of the AMSB example, it has only been shown
that $\mttwo(m_\ntlinoone)$ is bounded above by $m_\chginooneplus$.
It has not yet been shown that $\mttwo$ can attain this bound.  The
purpose of this section is to show that it can.}  We start from
definition \myeqref{MT2:MT2DEF}.  We also describe the region of
decay-phase-space which contains events which occur close to this
kinematic endpoint.

\begin{figure}[t]\begin{center}
\epsfig{file=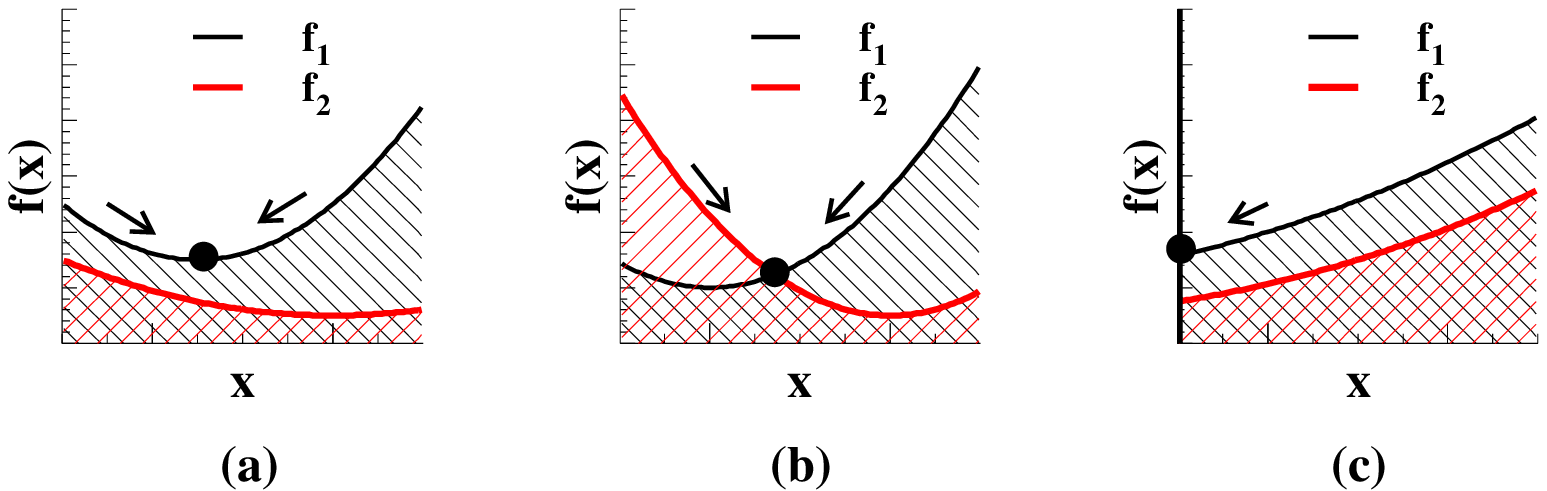, width=15cm}
\capbox{}{A diagram demonstrating that the minimisation over some parameter of the maximum of two well-behaved functions may occur either at {\bf (a)} a minimum value of one of them, or {\bf (b)} when they are equal,
or {\bf (c)} at the boundary of the domain.}
\label{MTX:MINIMISING}
\end{center}\end{figure}

\par

To find the range of values $\mttwo$ may take we first
let $f_1=m_T^2({\bf p}_T^{\pi^{(1)}}, \slashchar{{\bf q}}_T^{(1)}; m_\ntlinoone)$, 
and $f_2=m_T^2({\bf p}_T^{\pi^{(2)}}, \slashchar{{\bf q}}_T^{(2)}; m_\ntlinoone)$.
We then note that the minimum over a parameter $x$ 
of the maximum of $f_1(x)$ and $f_2(x)$ can occur at a local minimum, $f_{1(2)}^\prime(x^*)=0$, 
provided $f_{1(2)}(x^*)>f_{2(1)}(x^*)$, as shown in \myfigref{MTX:MINIMISING}a. 
Alternatively the minimum can occur when the functions cross one another
when $f_1=f_2$ (\myfigref{MTX:MINIMISING}b) or at a boundary (\myfigref{MTX:MINIMISING}c).
The parameter $x$ corresponds to the fraction of the the missing momentum 
(in one of the transverse directions) which is assigned to each half of the event.
Since $f_1, f_2\to\infty$ as $x\to\pm\infty$ \myfigref{MTX:MINIMISING}c is not relevant to our minimisation problem. 

\par

To find which of (a) or (b) is pertinent, consider an {\em unconstrained}
minimisation over $\slashchar{{\bf q}}_T$, of $m_T^2 ( {\bf
p}^{\pi}_T, \slashchar{{\bf q}}_T; m_\ntlinoone)$.  Using the relationship
\begin{equation}\frac{\partial \slashchar{E}_T}{\partial \slashchar{q}_k}=\frac{\slashchar{q}_k}{\slashchar{E}_T}\ ,
\end{equation}
where $\slashchar{E}_T^2=\slashchar{{\bf q}}_T^2+m_\ntlinoone^2$, it is straightforward to show that,
\begin{equation}\frac{\partial m_T^2}{\partial \slashchar{q}_k}=2\left(E_T^\pi\frac{\slashchar{q}_k}
{\slashchar{E}_T}-p_k^\pi\right) \qquad k=1,2\ .\end{equation} This
means that at an unconstrained minimum of $m_T^2$ we have
\begin{equation} {\bf v}_T^\pi = \slashchar{{\bf u}}_T\ , \label{MTX:GLOBMIN}\end{equation}
where we introduce the notation ${\bf v}_T \equiv {\bf p}_T / E_T$,
$\slashchar{{\bf u}}_T \equiv \slashchar{{\bf q}}_T /
\slashchar{E}_T$, in which ${\bf p}_T$ and ${\bf v}_T$ represent the true
transverse momentum and velocity of a particle, while $\slashchar{{\bf
q}}_T$ and $\slashchar{{\bf u}}_T$ are assigned by the minimisation.

\par

Using the basis ($t$,~$x$,~$y$) with the metric diag(1,-1,-1), one can write
\begin{equation} m_T^2 = (E_T^\mathrm{tot},{\bf p}_T^\mathrm{tot})\cdot(E_T^\mathrm{tot},{\bf p}_T^\mathrm{tot})\ ,
\label{MTX:MTLOR2} \end{equation} 
where $E_T^\mathrm{tot}=E_T^\pi+\slashchar{E}_T$ and 
${\bf p}_T^\mathrm{tot}={\bf p}_T^\pi+\slashchar{{\bf q}}_T$. 
This 1+2 dimensional Lorentz invariant can be evaluated in any frame boosted from the lab 
in the transverse plane. 
\myeqref{MTX:GLOBMIN} has told us that at the unconstrained minimum the
transverse velocities ${\bf v}_T^\pi$ and $\slashchar{{\bf u}}_T$ are equal;
a statement necessarily true in all transverse frames, including the special one
in which both the transverse velocities (and associated momenta) are zero.
Evaluating \myeqref{MTX:MTLOR2} in this frame, we find that the unconstrained minimum
%Now at the minimum \myeqref{MTX:GLOBMIN} will apply in the lab frame, 
%so it will always be possible to carry out a transverse boost into a frame 
%where ${\bf v}_T^\pi = \slashchar{{\bf u}}_T=(0,0)$.
%In the boosted frame the RHS 
of \myeqref{MTX:MTLOR2} then becomes $(m_\pi+m_\ntlinoone,~0,~0)\cdot(m_\pi+m_\ntlinoone,~0,~0)$, 
and we recover the expected result \begin{equation} m_T^\amin = m_\pi+m_\ntlinoone\ .\label{MTX:MTMIN} \end{equation}
We therefore conclude that the function $m_T^2$ has only one stationary value and it is the global minimum, 
and is common to both sides of the event provided the same type of particles are emitted.
Thus when $f_1$ is minimum it cannot be greater than $f_2$, 
and so the minimisation in \myeqref{MT2:MT2DEF} forces $f_1=f_2$.
This could of course occur when {\em both} $f_1$ and $f_2$ are at their global minima, 
in which case $\mttwo$ takes its minimum value:
\begin{equation}\mttwo^\amin=m_\pi+m_\ntlinoone\label{MT2:MT2MIN}\ .\end{equation}

\par

To summarise, when the same particles are emitted from both sides of
the event, $\mttwo$ may be defined as the minimum of $m_T^{(1)}$
subject to the two constraints $m_T^{(1)}=m_T^{(2)}$, and ${\bf
\pmiss}_T ^{(1)} + {\bf \pmiss}_T ^{(2)} = {\bf \pmiss}_T$.  The
condition for the minimisation can be calculated by
lagrange-multiplier methods, the result of which is that the velocity
vectors ${\bf \slashchar{u}}_T^{(1,2)}$ of the {\em assigned}
neutralino momenta $\slashchar{{\bf q}}_T^{(1,2)}$ must satisfy
\begin{equation}
({\bf \slashchar{u}}_T^{(1)}-{\bf v}_T^{\pi^{(1)}})\ \propto\ ({\bf \slashchar{u}}_T^{(2)}-{\bf v}_T^{\pi^{(2)}})\ .
\label{MTX:MTTWOCOND} \end{equation}

To find the maximum of $\mttwo$ over many events
we note that for each event the minimisation will select hypothesised momenta satisfying \myeqref{MTX:MTTWOCOND}.
We now note events can occur in which the {\em true} transverse velocities of the neutralinos 
were exactly those which were assigned by the minimisation, \ie\ they can satisfy 
\begin{equation}
{\bf v}_T^{\ntlinoone(1)} = {\bf \slashchar{u}}_T^{(1)}, \qquad {\bf v}_T^{\ntlinoone(2)} = {\bf \slashchar{u}}_T^{(2)}\ .
\label{MT2:TRUEVEL}
\end{equation}
These events will have both hypothesised transverse masses equal not only to each other but also to 
true transverse masses which would have been calculated if the neutralino momenta had been known:
\begin{equation}
m_T^{(i)} \left( {\bf p}^{\pi^{(i)}}_T, \slashchar{{\bf p}}_T^{\ntlinoone(i)} \right)=
m_T^{(i)} \left( {\bf p}^{\pi^{(i)}}_T, \slashchar{{\bf q}}_T^{(i)}\right) 
\end{equation}
%This means that with a large number of events there will be {\em some} events where the maximum limit that can be put 
%on the transverse masses will be that event's true transverse mass:
%\begin{equation}
%\mttwo=m_T^{(1)} ( {\bf p}^{\pi_1}_T, \slashchar{{\bf q}}_T^{(1)}) = 
%m_T^{(2)} ( {\bf p}^{\pi_2}_T, \slashchar{{\bf q}}_T^{(2)}) = 
%m_T^2 ( {\bf p}^{\pi_1}_T, \slashchar{{\bf p}}_T^{\chi_1}) =
%m_T^2 ( {\bf p}^{\pi_2}_T, \slashchar{{\bf p}}_T^{\chi_2})\ .
%\end{equation}
If events occur where, in addition to the transverse components of the neutralino momenta satisfying \myeqref{MT2:TRUEVEL}, 
the rapidity differences satisfy $\eta_{\ntlinoone(1)}=\eta_{\pi(1)}$ and  $\eta_{\ntlinoone(2)}=\eta_{\pi(2)}$,
then by \myeqref{MT2:MCHIDEF} $\mttwo$ will equal the true mass of the chargino.
Combining this with \myeqref{MT2:MT2MIN} and recalling that \label{sec:secinwhichmt2rangeforidentdecaysisdetailed}
$\mttwo$ cannot be greater than the chargino mass by construction,
we can see that the event-by event distribution of $\mttwo$ can span the range
\begin{equation}
m_\ntlinoone+m_\pi\ \leq\ \mttwo\ \leq\ m_\chginooneplus.
\end{equation}

%The variable is equally applicable to two 
% same-sign $\chginooneplus$ decays so $\mttwo$ 
%signal events can be 
%defined as those having two $\cht_1^\pm \to 
%\ntlinoone\ \pi^{\pm}$ decays with any 
%combination of charges.

\subsection{Extremal values of $\mtx$}

In the last section we looked at the conditions under which events can
generate $\mttwo$ values near the kinematic endpoint.  Here we will
look at some of the ways these conditions become modified for
$\mtthree$ and $\mtfour$ events.

\par

Consider once again events from the AMSB scenario in which a chargino
is produced and then decays to $\ntlinoone\ e\ \nu_e$.  If we expand
the Lorentz invariant\begin{equation} (m_\chginooneplus)^2 =
(p_\ntlinoone + p_e + p_\nu)^2
\end{equation}
we obtain three mass-squared terms for each of the decay particles and three cross-terms. 
The cross-terms can each be written in the form
\begin{equation}
2 p_a \cdot p_b = 2 \left[ { E_T^{(a)}E_T^{(b)}\cosh(\Delta\eta_{ab})-{\bf p}_T^{(a)}\cdot{\bf p}_T^{(b)}} \right],
\end{equation}
like the cross term in \myeqref{MT2:MCHIDEF}.  If the neutralino and
neutrino transverse momenta were individually known we could evaluate
the transverse mass,
\begin{equation} m_T^2 = m_\ntlinoone^2 + m_e^2 +
2\ \Bigl[ (E_T^e E_T^\chi - {\bf p}_T^e \cdot {\bf p}_T^\chi ) + 
(E_T^\nu E_T^\chi - {\bf p}_T^\nu \cdot {\bf p}_T^\chi ) + 
(E_T^e E_T^\nu - {\bf p}_T^e \cdot {\bf p}_T^\nu ) \Bigr]\ ,\label{MTX:MT3PART}\end{equation}
where the neutrino mass is assumed to be negligible.
$m_T$ will be equal to the $\chginooneplus$ mass in events where $\Delta\eta_{ab}=0$ for all pairs of 
$e$, $\nu_e$, and $\ntlinoone$.

\par

Using, in \myeqref{MT2:MT2DEF}, the three-particle definition of $m_T$
from \myeqref{MTX:MT3PART} instead of the two-particle definition
\myeqref{MT2:MTDEF}, one defines $\mtfour$, the analogue of $\mttwo$
for the case of four missing particles.  The constraint on the
unobserved momenta will, of course, have to be modified to read
\begin{equation}
{\bf q}_T^{\nu(1)} + {\bf q}_T^{\chi(1)} +{\bf q}_T^{\nu(2)} + {\bf
q}_T^{\chi(2)} = \Ptmiss\ ,
\end{equation}
where the labels (1) and (2) indicate which chargino the particles
were emitted from.

\par

The conditions for the minimisation required to calculate $\mtfour$ can
be calculated just as for $\mttwo$.  The Euler-Lagrange (E-L) equations
involving
\begin{equation}
\frac{\partial (m_T^{(i)})^2}{ \partial {\bf q}_T^{\nu(i)} } \qquad
\mathrm{and} \qquad \frac{\partial (m_T^{(i)})^2}{\partial {\bf
q}_T^{\ntlinoone(i)} }\nonumber
\end{equation} show that the minimisation will select the
invisible particles' momenta such that ${\bf u}_T^{\ntlinoone(i)}={\bf
u}_T^{\nu(i)}$.  The other E-L equations reproduce
\myeqref{MTX:MTTWOCOND} but with electrons replacing pions.

\par

This means that when calculating $\mtfour$ one can replace the missing
particles from each chargino decay with a pseudo-particle with mass
equal to the sum of the masses of those invisible particles and
proceed as for $\mttwo$.  In the case of leptonic chargino decay the
mass of the neutrino can be safely neglected in comparison to that of
the $\ntlinoone$, and the constraint ${\bf u}_T^{\chi(i)}={\bf
u}_T^{\nu(i)}$ is equivalent to ${\bf q}^{\nu(i)}_T=(0,0)$.

\par

\label{sec:numbersofeventsnearthetails}

The distribution over events of $\mtfour$ will have fewer entries near
the upper kinematic limit ($\mtfour=m_\chginooneplus$) because when
more particles go undetected an event at that limit must satisfy a
larger number of constraints.  For fully leptonic chargino-pair decay,
there are six constraints of the type $\Delta\eta=0$, two ${\bf
p}_T^{\nu(i)}=0$ and finally the modified constraint from
\myeqref{MTX:MTTWOCOND}.  This effect can be seen in
\myfigref{fig:deltamtxhisto} for events where a total of two, three and
four invisible particles are produced.

\subsection{Asymmetric decays}

In the preceding two sections we have seen that when the decays on
each side of the event are the same (\ie\ both initial sparticles
decay to the same set of daughter particles) then the resulting
kinematic variables, $\mttwo$ and $\mtfour$, have very similar
properties.  The only significant difference we have seen is the
reduced density of events near the upper kinematic endpoint of
$\mtfour$ relative to $\mttwo$.  Why, then, is the $\mtthree$
distribution, shown in \myfigref{fig:deltamtxhisto}, seen to have a
shape significantly different from the $\mttwo$ and $\mtfour$
distributions?  Specifically, why does it have the strong peak at low
values not shared by the other two?

\par

The difference occurs because the visible particles on each side of an
$\mtthree$ event are different (on one side $\ntlinoone,e,\nu$ and on
the other to $\ntlinoone,{\pi^+}$) and so the unconstrained minima
of the values of $m_T$ on each side of the event are not equal as they
are in the case of $\mttwo$ and $\mtfour$:
\label{sec:dfgdkferhg}
\begin{equation}
\min_{\slashchar{\bf q}_T^{(1)}} \left(m_T^{(1)}( {\bf p}^{\pi}_T, \slashchar{{\bf q}}_T^{(1)})\right) 
\ =\ m_\pi + m_\ntlinoone\quad \ne\quad m_e + m_\ntlinoone\ =\ 
\min_{\slashchar{\bf q}_T^{(2)}} \left(m_T^{(2)}( {\bf p}^{e}_T, \slashchar{{\bf q}}_T^{(2)})\right)
\end{equation}
It is thus possible for some of the events can then fall into the
category shown in \myfigref{MTX:MINIMISING}a, producing a peak of
events with $m_{TX}=m_\ntlinoone+m_\pi$.

\section{LHC case studies using $\mttwo$}
\label{sec:realmt2}
%%%%%%%%%%%%%%%%%%%%%%%%%%%%%%%%%%%%%%%%%%%%%%
In this section we highlight some physics studies for the LHC
which demonstrate that the background and the detector effects do not 
prevent $\mttwo$ from being a useful experimental variable. 
We investigate points from three different models,
under two different classes of mass hierarchy.

\subsection{Case 1 -- mSUGRA-like points}
\label{mSUGRA}
The first two points discussed are the mSUGRA point 5 (S5) 
and a point from the optimised string model
(O1) discussed in \cite{LESTER:THESIS}. The relevant parameters of these 
models are
\begin{eqnarray}
&m_{3/2}& = 300 ~\GeV, m_0 = 100 ~\GeV, A_0 = 300 ~\GeV, \tan \beta = 2.1, 
\nonumber \\
&m_{3/2}& = 250 ~\GeV, \tan \beta = 10, \theta = \pi / 4 \nonumber 
\end{eqnarray}
respectively and $\mu > 0$ in both cases. For these points we are looking at
dislepton production from a hard process which decays as $\tilde{l}^\pm 
\rightarrow \tilde{\chi}^0_1 l^\pm$, and so the 
mass-hierarchy is $m_l \ll m_{\tilde{l}}-\chi \approx \chi < m_{\tilde{l}}$.

For S5 and O1, all events, except the $qq \rightarrow W^+ W^-$ background 
processes, were simulated by {\tt HERWIG-6.0}~\cite{herwig61}. The 
W-pair events were generated by {\tt ISAJET-7.42}~\cite{isajet740}. 
The events for these two points were generated at 100 ${\rm fb}^{-1}$. This is
expected to correspond to running at high luminosity for one year.

\begin{figure}[t]
\begin{center}
\subfigure[S5, hard cuts]{
\epsfig{file=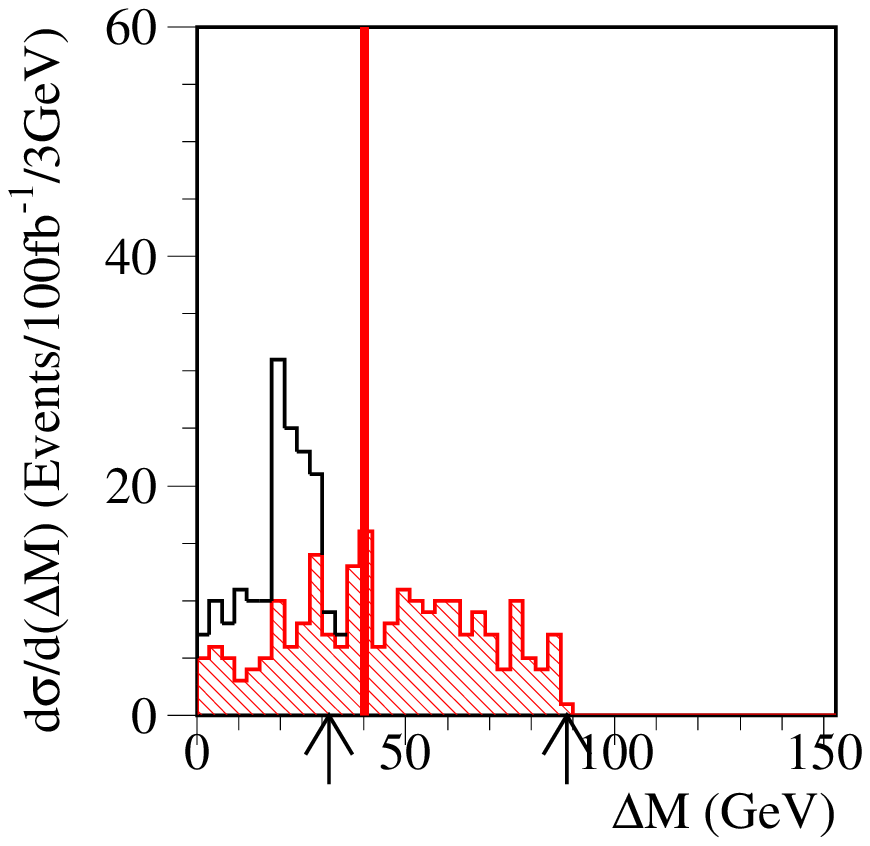,width=0.45\textwidth} }
\subfigure[O1, hard cuts]{
\epsfig{file=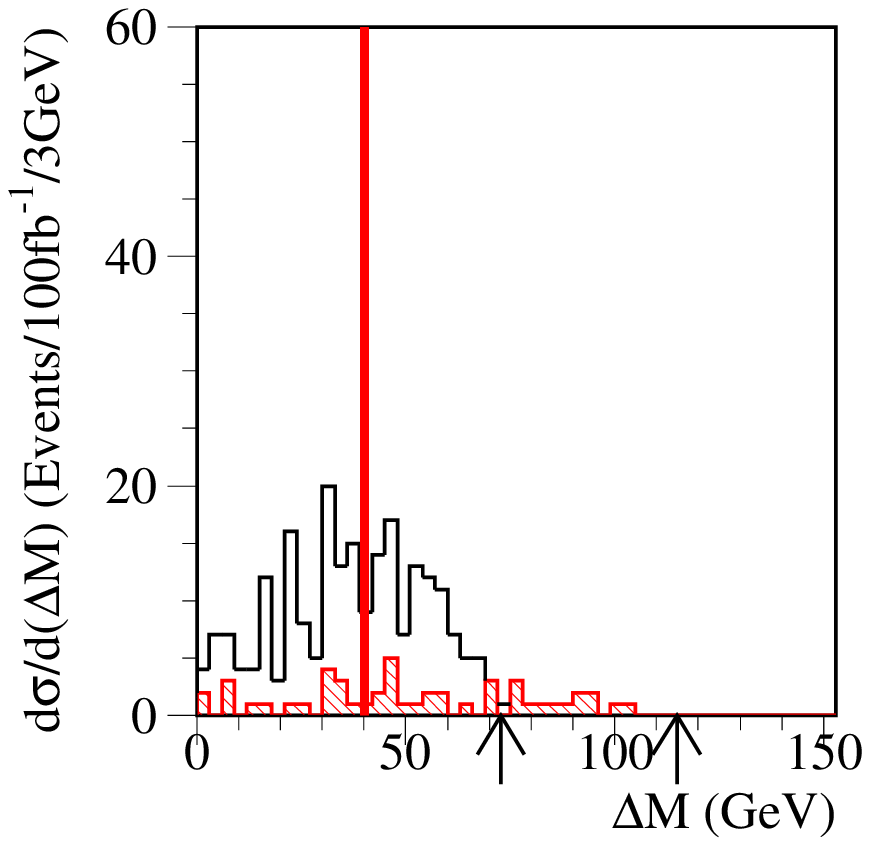,width=0.45\textwidth} }
\capbox{Signal dislepton events}{$\Delta M$ distributions obtained
at S5 and O1 after applying the cuts to a $100~\ifb$ sample of ${\tilde e}_R
{\bar{\tilde e}}_R$, ${\tilde \mu}_R {\bar{\tilde \mu}}_R$, ${\tilde
e}_L {\bar{\tilde e}}_L$ and ${\tilde \mu}_L {\bar{\tilde \mu}}_L$
dislepton events.  Events from light right-sleptons (unhatched) are
stacked on top of those from heavier left-sleptons (hatched).  Only
events from OSSF leptons combinations are shown.  The plots are
generated without OSDF background subtraction, but were it to be
performed, no significant differences would be apparent as only 4 (12)
signal events are able to pass OSDF soft cuts at S5 (O1).
Arrows indicate the values of $\Delta M^{\rm max}$ predicted by theory for
the two types of slepton in each model.  A red vertical line is drawn
through each plot at half the $W$ mass.\label{fig:s5_o1_mt2_SIG}}
\end{center} \end{figure}

Since there are two different processes being analysed, there are different
cuts to apply. As this is not intended to introduce new physics, here we 
present only the major cuts used. For more detail about the cuts and the 
techniques used, see~\cite{LESTER:THESIS, Barr:2002ex}.

The events used for S5 and O1 are required to have one opposite sign same 
family (OSSF) pair of isolated leptons with $p_T^{l_1} > 50~\GeV$ and 
$p_T^{l_2} > 30~\GeV$. These events cannot contain any other isolated leptons.
Also, events containing one ore more jets with $p_T^j > 40~\GeV$ are vetoed. 
This helps reduce the standard model backgrounds.

The variable $\Delta M$ is defined as
\begin{equation}
\left( \Delta M \right) \equiv \frac{1}{4} \left( M_{T2}^2(m_l)\right)^2 
- m_l^2.
\end{equation}
This variable is what is studied for the points S5 and O1, for reasons given in
\cite{LESTER:THESIS}. The desired dislepton events have very little jet 
activity and the dislepton production cross sections are typically two orders 
of magnitude smaller than the squark/gluino production cross sections.
There are also irreducible SM backgrounds (primarily
$W^+W^-\rightarrow l^+ l^- \nu \bar \nu$ and $\ttbar\rightarrow b \bar
b W^+W^- \rightarrow j j l^+ l^- \nu \bar \nu$ in cases where jets are
below the reconstruction threshold or are outside detector acceptance)
which have signatures identical to dislepton events.  The smallness of
the signal and the presence of these backgrounds would cause problems
for naive straight-line fitting technique. Instead, the technique described
in~\cite{LESTER:THESIS} is used for the estimation of the edge precision. 

\begin{figure}[t]
\begin{center}
\subfigure[S5]{
\epsfig{file=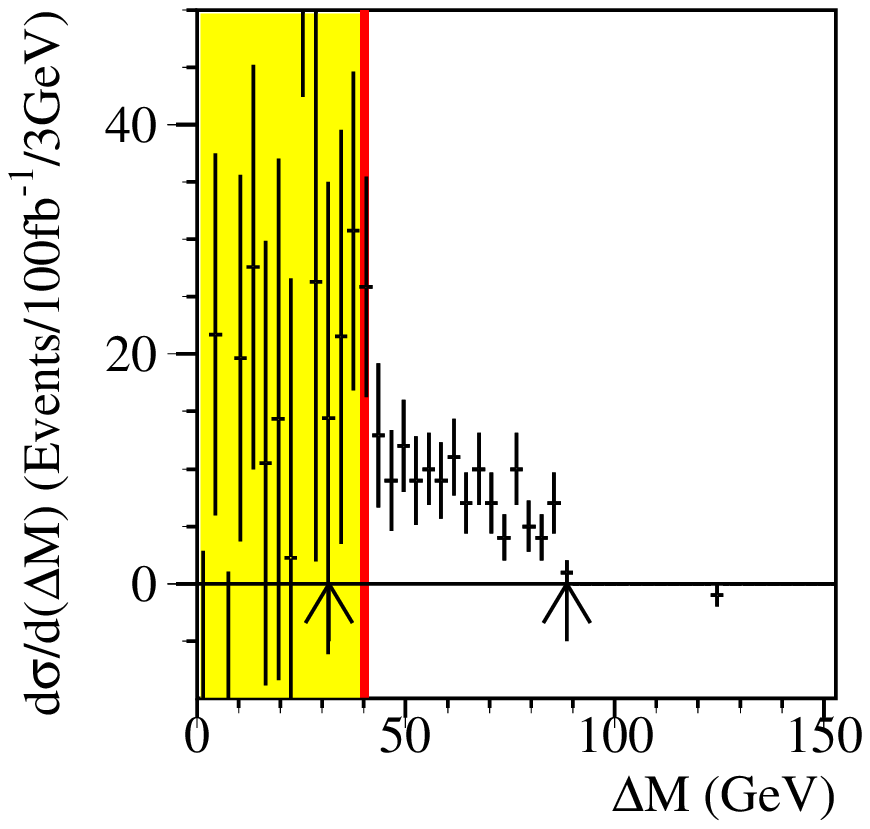,width=0.45\textwidth} }
\subfigure[O1]{
\epsfig{file=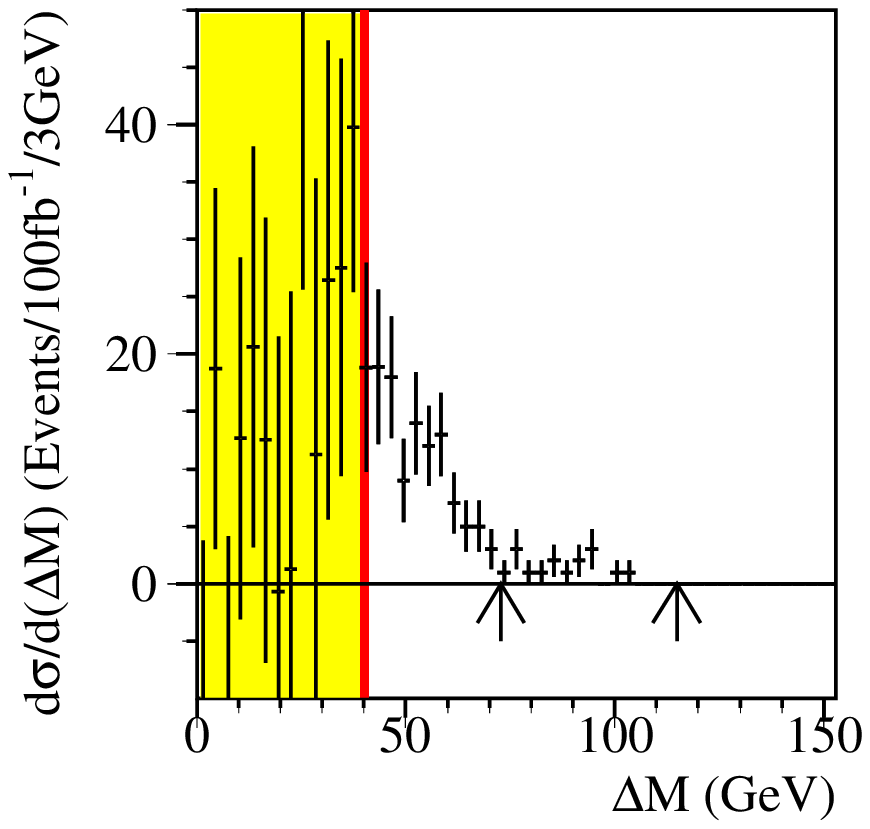,width=0.45\textwidth} }
\capbox{$\Delta M$ distributions}{Flavour-subtracted
$\Delta M$ distributions for combined signal and background at S5
and O1 after applying the cuts.  A red vertical line is drawn through each
plot at half the $W$ mass.  Plots are shaded to the left of this line
in order to draw attention to the events which reconstruct {\em above}
this point.  Compare these plots with those of
Figure~\ref{fig:s5_o1_mt2_SIG} which contain only signal events.  The
arrows of Figure~\ref{fig:s5_o1_mt2_SIG} are provided for comparison.
\label{fig:s5_o1}} \end{center} \end{figure}

Figure~\ref{fig:s5_o1_mt2_SIG} shows the $\Delta M$ distributions
obtained at S5 and O1 after applying the cuts to a $100~\ifb$ sample of signal 
dislepton events.  Events from lighter sleptons (${\tilde e}_R$ and ${\tilde
\mu}_R$) occupy the unhatched region in each plot, while the events
from heavier left-sleptons are cross hatched.  It will be noted that
events from both light and heavy sleptons succeed in passing the cuts
in both models.  In principle, then, there are two edges to be
observed in each of the models: one for the lighter slepton and one
for the heavier.  We note, however, that as the slepton masses
increase, their production is strongly suppressed, and so there are
very few heavy slepton events at O1 where there are in fact none
within $10~\GeV$ of the kinematic limit.  It is readily observed that
at the three remaining edges, where statistics are higher, there is
good agreement between the theoretical prediction and the observed
endpoint of each distribution.
 
Significant numbers of SM background events also pass the cuts. These can be
well modelled by its opposite sign different family (OSDF) counterpart. As
the signals from dislepton pair production are expected to be purely OSSF,
we can use OSDF background subtraction. Supersymmetric backgrounds also have
to be considered. Again, in this case the OSSF distributions are well modelled
by OSDF events passing the same cuts. So again, different-family background
subtraction is used.

All events for S5 and O1 (signals and backgrounds) are combined in
Figure~\ref{fig:s5_o1} after different-family background
subtraction.  The reader is encouraged to compare these plots with
those from Figure~\ref{fig:s5_o1_mt2_SIG} showing the desired signal
shapes. As expected, all signal shape information is lost to the left
of $m_W/2$ due to obliteration by the SM backgrounds.  To the right
of this point, at least one clear edge is observable in both models
(the left-slepton edge at S5, and the right-slepton edge at O1)
and in both sets of cuts.  The hard cuts are able to suppress the
supersymmetric backgrounds to such a degree that there is even compelling
evidence at O1 for the existence of {\em two} edges, although the
lack of statistics in the higher edge limits the precision with which
the endpoint may be located.

\subsection{Case 2 -- AMSB-like scenarios}
\label{AMSB}

\begin{table}\begin{center}
\begin{tabular}{|l|c|c|c|c|}
\hline  & $m_\chgone$ & \DeltaMChi &  & \\
\raisebox{1.5ex}[0pt]{Point}   &   (GeV)    & (MeV) &  
%\raisebox{1.5ex}[0pt]{$\chgone\to\ntlone \pi^+$ } &
\raisebox{1.5ex}[0pt]{$\chgone\to\ntlone e^+\nu_e$ } & 
\raisebox{1.5ex}[0pt]{$\chgone\to\ntlone \mu^+\nu_\mu$ }
\\\hline
{\bf SPS-300} & 165 & 886 & 17.0 \% & 15.9 \% \\
{\bf A-250}   & 101 & 766 & 15.4 \% & 13.9 \% \\
\hline
{\bf SPS-250} & 159 & 1798 & 21.9 \% & 21.5 \% \\
{\bf A-200}   & 97  & 1603 & 22.5 \% & 22.2 \% \\ \hline
\end{tabular}
\capbox{}{The lightest chargino mass, the mass difference, 
$\DeltaMChi=m_\chgone-m_\ntlone$, and two chargino branching ratios for the
AMSB-like points discussed in \mysecref{AMSB}.
The hadronic branching ratios can be found in \cite{Barr:2002ex}.
\label{tab:amsb}}
\end{center}\end{table}

The characteristic signature for anomaly-mediated supersymmetry breaking
is the near mass-degeneracy of the lightest chargino and the lightest
neutralino. The $\chgone$ therefore decays to a neutralino plus (relatively) light
standard-model particles. For a small mass difference, 
$\DeltaMChi=m_\chgone-m_\ntlone$,
the largest $\chgone$ branching ratios are to $\ntlone \pi^+$ and to $\ntlone l^+ \nu_l$,
where $l\in e,\mu$.
The mass hierarchy,
\[m_\pi\ \mathrm{or}\ (m_l + m_\nu) \approx m_\chginooneplus-\chi \ll \chi \leq m_\chginooneplus\ ,\]
is therefore very different to the previous case study.

{\tt HERWIG-6.3} was used to generate 30 ${\rm 
fb}^{-1}$ of unweighted inclusive supersymmetry events. 
{\tt HERWIG} was also used to generate the
background. For all the points, the results were passed through the ATLAS 
fast detector simulator, {\tt ATLFAST}\cite{atlfast20}.
The signal-enhancing cuts require missing transverse energy, 
$\slashchar{E}_T^{\rm min} = 500~\GeV$, leading jet transverse momentum, 
$p_{T(J_1)}^{\rm min} = 400~\GeV$ and transverse sphericity, $S_T^{\rm min} 
= 0.05$. There are also cuts on the tracks, these are described in more detail
in~\cite{Barr:2002ex}.

We consider AMSB-like points, which have the following parameters:
\begin{eqnarray}
&m_0& = 450 ~\GeV, m_{3/2} = 60~{\rm TeV}, \tan \beta = 10, \mu > 0,\nonumber\\
&m_0& = 500 ~\GeV, m_{3/2} = 36~{\rm TeV}, \tan \beta = 10, \mu > 0. \nonumber
\end{eqnarray}
and for which the $\mu$ parameter has been adjusted at the electroweak
scale in order to investigate 
different values of $\DeltaMChi$, as discussed in \cite{Barr:2002ex}.
Some masses and branching ratios can be found in \mytabref{tab:amsb}.

\begin{figure}[tb]
\begin{center}
\subfigure[SPS-300]{
\epsfig{file=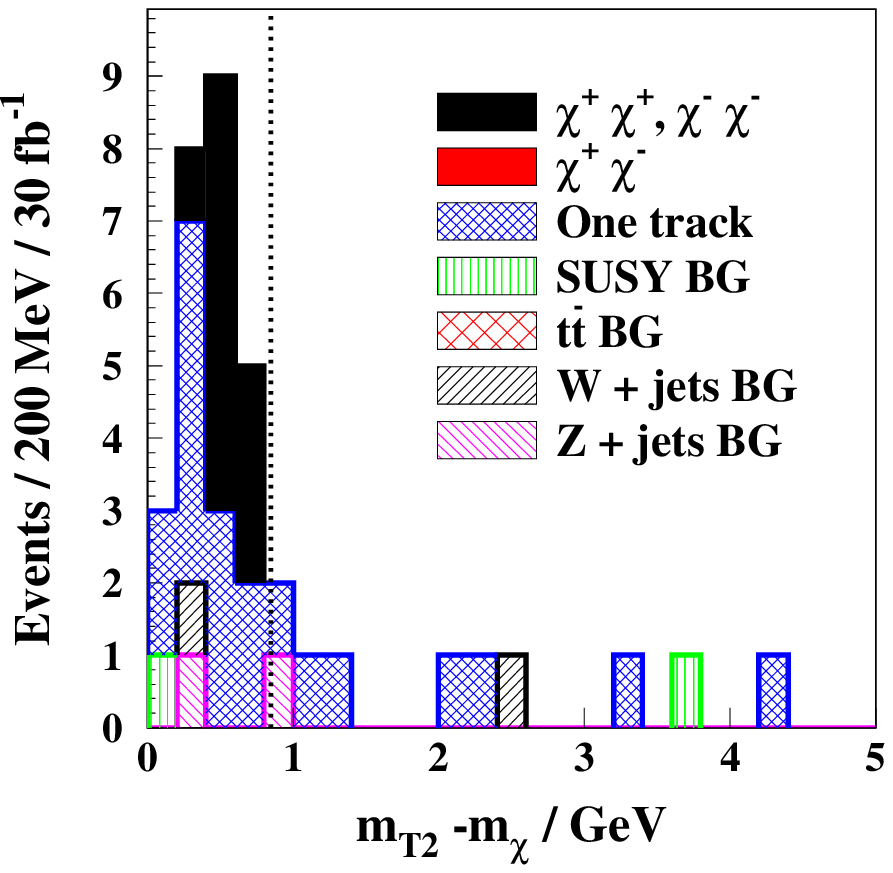,width=0.45\textwidth} }
\subfigure[A-250]{
\epsfig{file=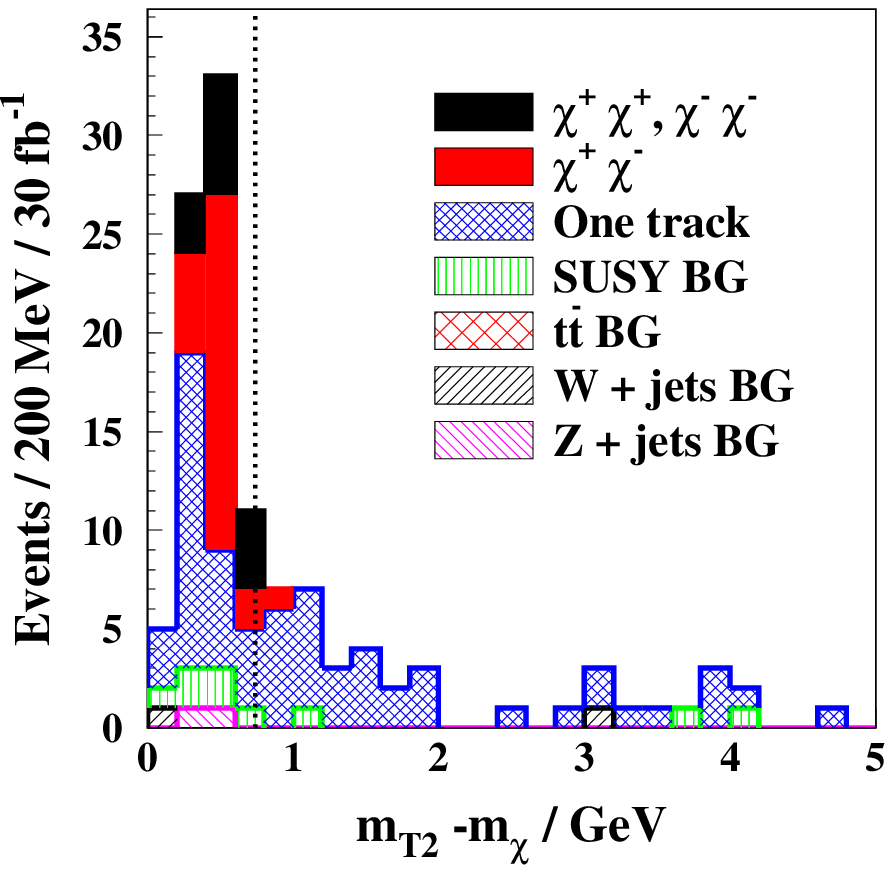,width=0.45\textwidth} }
\capbox{}{The $\mttwo-m_{\tilde{\chi}^0_1}$ distribution for {\bf (a)} the point 
{\bf SPS-300}, and {\bf (b)} the point {\bf A-250}.
The signal consists of the two solid regions labelled $\chi^x+\chi^x$ in the legend.
The upper kinematic limit of $\mttwo-m_\ntlone$
for signal events is marked with a dotted line.
Note the sharp fall-off in the distribution near the kinematic edge
at $\mttwo-m_\ntlone=\DeltaMChi$.
\label{fig:sps300_a250}} \end{center} \end{figure}

\begin{figure}[tb]
\begin{center}
\subfigure[SPS-250]{
\epsfig{file=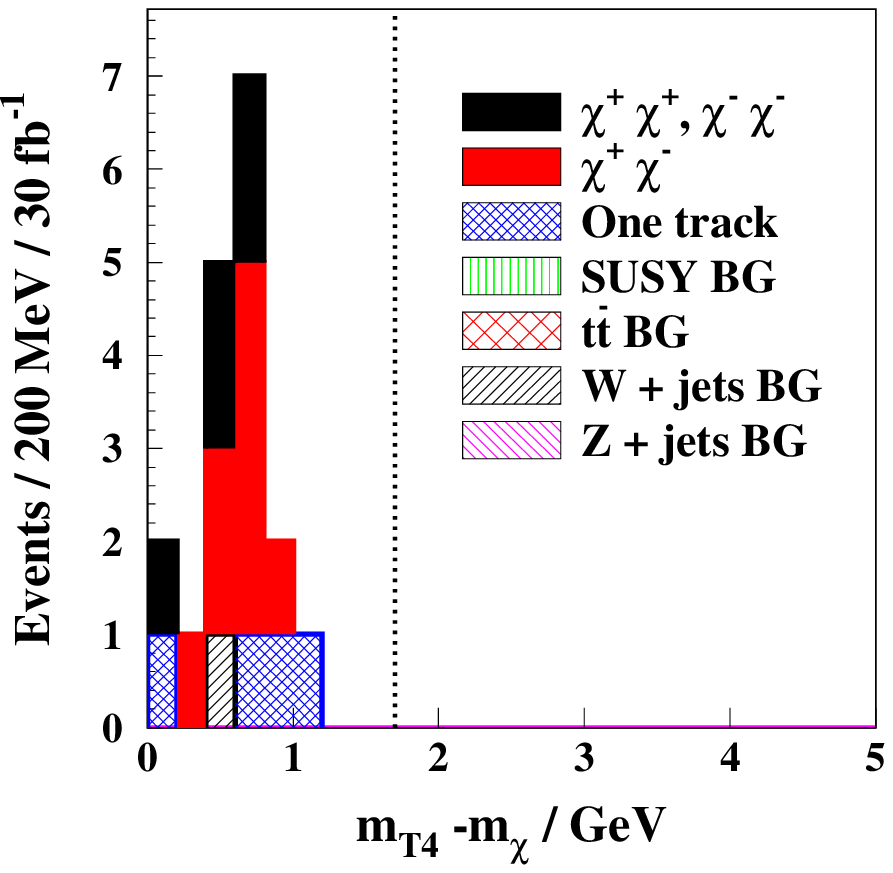,width=0.45\textwidth} }
\subfigure[A-200]{
\epsfig{file=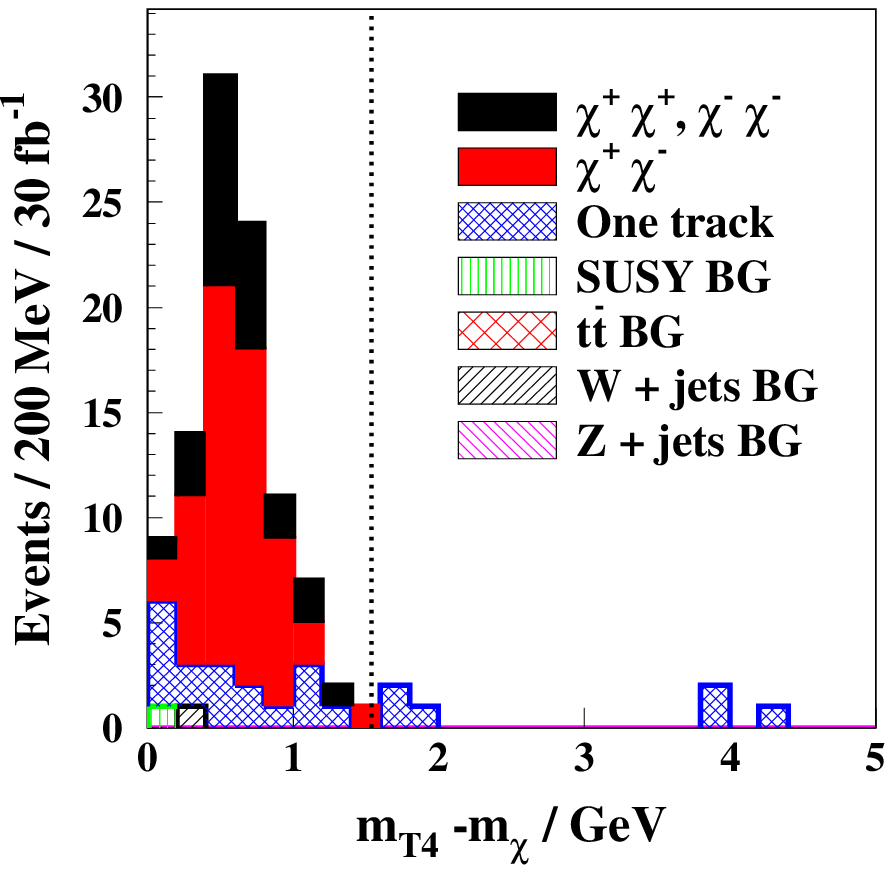,width=0.45\textwidth} }
\capbox{}{The $\mttwo-m_{\tilde{\chi}^0_1}$ distribution for {\bf (a)} the point 
{\bf SPS-250}, and {\bf (b)} the point {\bf A-200}.
The signal consists of the two solid regions labelled $\chi^x+\chi^x$ in the legend.
The upper kinematic limit of $\mttwo-m_\ntlone$
for signal events is marked with a dotted line.
Note the fall-off in the distribution near the kinematic edge
at $\mttwo-m_\ntlone=\DeltaMChi$.
\label{fig:sps250_a200}} \end{center} \end{figure}

The first two points have large branching ratios for the decay 
$\tilde{\chi}^\pm_1 \rightarrow \pi^\pm \tilde{\chi}^0_1$.
This means that chargino-pair decay can easily generate the
topology shown in \mysecref{fig:schematic}.
We therefore plot distributions of $\mttwo - m_{\tilde{\chi}^0_1}$, 
for which signal events in a perfect detector would lie in the range 
$[m_\pi, \DeltaMChi]$.
The results (see figure \ref{fig:sps300_a250}) show that 
$\mttwo$ could be used to measure the small mass difference between the 
$\tilde{\chi}^+_1$ and the $\tilde{\chi}^0_1$ in this model, provided the 
signal cross-section is sufficiently large.

The second pair of points each have a larger leptonic 
branching ratio, and so for these points
the fully leptonic channel was investigated.
Since there are now four missing particles in the final state,
(two neutralinos, and two neutrinos),
distributions of $\mtfour-m_\ntlone$ were plotted.
For a perfect detector, these
are restricted to lie in the range  $[m_{e/\mu}, \DeltaMChi]$.

The signal events are again indicated by the solid shades in the 
histograms in \myfigref{fig:sps250_a200}.
Again, it can be observed that the distribution lies within
the expected range. The distribution is 
skewed to lower values because more particles are missing, and so more 
constraints must be satisfied for an event to approach the 
upper limit (as was seen in \myfigref{fig:deltamtxhisto}).

The sensitivity of $\mtx$ to the estimated mass of the neutralino was
shown in \myfigref{MTX:MTXVERSUSCHIPLOTS}.  It has been found that
$\mtx$ shows similar insensitivity to measurement uncertainties in the
missing transverse momentum vector.  This behaviour can be (at least
partially) understood from the non-relativistic limit of $\mttwo$,
when the proportionality in \myeqref{MTX:MTTWOCOND} becomes an
equality and
\begin{equation} \mttwosq-(m_\pi+m_\ntlinoone)^2 = \frac{1}{4 m_\pi m_\ntlinoone}
\left(m_\pi \Ptmiss - m_\ntlinoone {\bf p}_T^{\pi_1} - m_\ntlinoone
{\bf p}_T^{\pi_2}\right)^2 + \mathcal{O}\left(({\bf v}_T\cdot {\bf
v}_T)^2\right)\ . \label{AMSB:NONREL}\end{equation} 
The low sensitivity to the (possibly poorly-measured) quantities 
$m_\ntlinoone$ and $\Ptmiss$ comes from the fact that 
in \myeqref{AMSB:NONREL}
they are multiplied by the quantities ${\bf p}_t^\pi$ and $m_\pi$ respectively,
which are both small in this mass regime.

\section{Conclusion}

This paper has attempted to achieve three objectives.  Firstly it
seeks to introduce a new set of kinematic variables $\{\mttwo$,
$\mtthree, ...\}$, which are specially designed to extract information
from a particular class of troublesome events that we are likely to
see at next generation hadron and lepton colliders.  These events are
those containing a pair of particles of identical (but unknown) mass
which subsequently decay into groups of particles, each containing one or
more invisible (possibly massive) particles.  An example of this kind
of event might be pair production of sleptons at the LHC, followed by
subsequent sleptonic decay to leptons and neutralinos.  Secondly this
paper attempts to get to the bottom of these new variables; it
describes the regimes with in which they can or cannot be trusted,
develops useful approximations to them, and shows generally how one
could go about calculating this variable for real.  The
approximations
to the variables are not only useful in their own right, but are even
more useful as guides which illustrate the dependence of the
variables
upon its inputs. Finally, this article seeks to show with a couple of
examples, real use of these variables in physics analyses.  These
hopefully show that $\mttwo$ and its chums are able to provide vital
and new information about particle masses from events that would at
first glance appear to contain so many unknown quantities as to be
useless.

\par

In conclusion, we believe that $\mttwo$ is invaluable tool
for physicists working at the LHC, and other future colliders,
and we hope that this document will encourage its use.

\bibliography{bib}

\end{document}